\documentclass[12pt]{elsarticle}

\usepackage{color}
\usepackage{amssymb}
\usepackage{fullpage,epsf,graphicx, amstext,url} 
\usepackage{natbib}
\usepackage{mathtools}
\usepackage[footnotesize]{caption}
\usepackage{float}
\usepackage[compact]{titlesec}
\usepackage{subfigure}
\usepackage{enumitem}
\usepackage{multirow}
\usepackage[bottom]{footmisc}
\usepackage{url}
\usepackage{caption}
\usepackage{array}
\usepackage{pdflscape}
\usepackage{picture}

\journal{Journal of Nuclear Materials}

\begin{document}

\begin{frontmatter}



\title{Mechanism for Radiation Damage Resistance in Yttrium Oxide Dispersion Strengthened Steels}


\author{ J. Brodrick, D.J.Hepburn and G.J.Ackland$^*$ }

\address{School of Physics, CSEC and SUPA, University of Edinburgh, JCMB, King's Buildings, Edinburgh EH9 3JZ, UK. $^*$Corresponding author: gjackland@ed.ac.uk +441316505299}

\begin{abstract}
  ODS steels based on yttrium oxide have been suggested as potential
  fusion reactor wall materials due to their observed radiation
  resistance properties. Presumably this radiation resistance can be
  related to the interaction of the particle with vacancies,
  self-interstitial atoms (SIAs) and other radiation damage debris.
  Density functional theory has been used to investigate this at the
  atomic scale. Four distinct interfaces, some based on HRTEM
  observations, between iron and yttrium oxide were investigated.  It
  is been shown that the Y$_2$O$_3$-Fe interface acts as a strong trap
  with long-range attraction for both interstitial and vacancy
  defects, allowing recombination without altering the interface
  structure.  The catalytic elimination of defects without change to
  the microstructure explains the improved behaviour of ODS steels with
  respect to  radiation creep and swelling.
\end{abstract}

\end{frontmatter}


\section{Introduction}
\label{Introduction}
Fusion energy has the potential to provide clean and abundant power
for future generations\citep{futureProspects}. Many designs have been
suggested to make this a reality: These include the
tokamak\citep{stellTok} and inertial confinement\citep{icf}. Current
facilities attempting to realise these designs are, respectively, the
International Thermonuclear Experimental Reactor (ITER) in
France, which is under construction, and the
National Ignition Facility (NIF) in California. Regardless
of design, any fusion reactor would have to withstand the intense
neutron radiation 
ITER is predicted 
to produce a flux rate of at least 1 MW m\textsuperscript{-2} at its
walls\citep{iterBasis}, equivalent to fifty thousand 14.1 MeV
neutrons incident per square centimetre every
nanosecond. 

Thus, it is critical to find materials that are resistant to neutron
radiation at elevated temperatures.   { There have been a large number of
other experiments on irradiated ODS
\citep{mechprops,radExp,review,swelling,Toloczko,Chen,Chen2,Kim,Zhang}
balance of experimental evidence suggests that ODS steels exhibit a
lower level of swelling and creep than equivalent steels without ODS
dispersoids.   For example, a strong candidate material for
the inner wall of a fusion reactor is the EUROFER97
ferritic/martensitic steel, a chromium-rich iron alloy. Oxide
dispersion strengthened steels\citep{odsSteels} based on EUROFER97 can
be constructed through a sequence of mechanical alloying and hot
isostatic pressing processes to incorporate yttrium oxide into the
metal\citep{decanano}. These ODS steels typically hold between between
0.3 and 0.5\% yttrium oxide by weight\citep{mechprops},  and are
remarkably durable, with the form of yttria grains remaining stable
after 590 MeV proton irradiation up to doses of 1.0 displacements per
atom (dpa)\citep{radExp}. These experiments have observed the
increased yield stress of ODS steels compared to the standard
EUROFER97 steel to be further enhanced under this level of irradiation.  }
Other improvements include increased ultimate tensile
strength, by a factor greater than 1.5, persisting up to 650$^\circ$C;
the creep strength at 700$^\circ$C is equal to that of the base steel
at 600$^\circ$C\citep{mechprops} and a 30-40\% greater yield
strength\citep{radExp}.

Some less ideal properties of the ODS steel include a reduced
elongation on fracture at high temperature and raised
ductile-to-brittle-transition temperature, rendering the ODS steel
brittle at room temperature, has been observed\citep{mechprops,radExp}.
However, both high temperature elongation and ductility can be
improved by the inclusion of titanium impurities\citep{radExp}.
These properties of ODS steel suggest that reactor
operating temperature could be increased by 100K or more, compared to
the base EUROFER97. 

Suggested reasons for the radiation resistance exhibited by ODS steels
have been summarised by Sch\"{a}ublin\citep{radExp}. One idea is that
the oxides provide a catalyst for annihilation of the structural
defects (vacancies and SIAs caused by radiation (see Section
2). This could either be due to some attraction between
the oxide and the defects, or  altered defect dynamics at the
interface boundary combined with the high density of particles.  
Another idea is that the disorder already brought
about by the oxide dispersions in the steel structure make further
disruptions caused by radiation less effective in weakening the
material.  A third is based on the original purpose of ODS: that the increased density
of pinned dislocations aids recombination.

We investigate the recombination-catalyst idea for the idealised 
case of ferritic iron and pure yttrium oxide (Y$_\mathrm{2}$O$_\mathrm{3})$. 
We use density functional theory (DFT)
simulations of possible interfaces between the oxide and the
metal. This allows for modelling of the behaviour of radiation induced
defects near the interface. In particular, the question to be answered
is whether these atomic scale methods will be able to provide an
explanation for the impressive performance of ODS steels under
irradiation. Additionally, it would be desirable if this explanation
would give a clue to improving these materials' properties.


 Pure iron has a simple body-centred cubic lattice structure up to
 1043 K, which contains the entirety of the operating
 temperature limits for a fusion reactor wall \citep{decanano}. A
 14.1MeV neutron penetrate deep into the metal and eventually collide
 with a lattice atom, this initiates a ``displacement cascade'' which
 creates vacancy and SIA defects in equal numbers.
\citep{radMatBook} 

The displacement cascade is a rapid process (of
order picoseconds). Further migration of vacancies and
SIAs, mainly by diffusion,  happens over a  timescale of order 
nanoseconds\citep{radMatBook}. 
This is
still short compared to operating times, so is important to consider
the equilibrium result of such processes: If the vacancies and
SIAs were likely to find their Frenkel partner, recombine,
and annihilate, then the metal should essentially return to its
original structure; however, if defects instead formed large clusters
of a single type this could result in formation of voids, dislocation
loops or swelling, possibly weakening the material in the
process.
Defects can be trapped at grain
boundaries or surface, so for an ODS particle to effect the diffusion, there concentration must be such that there are many such particles in each grain.

Atomic impurities dispersed throughout the metal play a significant
role in point defect dynamics\citep{radMatBook}. From elastic
considerations, one might expect that substitutional impurities would
attract SIAs, and oversized substitutional and interstitial defect
attract vacancies.  However, the stable SIA configuration iron is the
(110) dumbbell, which has a quadrupole strain field with both tensile
and compressive regions - thus the SIA binds to most impurities.
Moreover, the magnetic structure of iron can lead to complex binding
where the idea of ``big'' or ``small'' atoms is oversimplistic - for 
example Cr behaves like an undersized defect interacting with defects, 
but when added to Fe increases the lattice parameter; Ni is exactly opposite\citep{FeNiCr}.  Recent DFT simulations\citep{gopejenko,PERFECT,FeTM,Murali},
in iron lattices, have shown well defined trends across the periodic table for defect binding in both fcc and bcc iron.

\section{Yttrium Oxide in Iron}

\subsection{Pure Yttrium Oxide}
The structure of pure yttrium oxide, $\mathrm{Y_2O_3}$, is described
by the $Ia\bar{3}\ (T^7_h)$ space group,\citep{Y2O3} which has
body-centred symmetry. This structure can be represented by an 80 atom
cubic unit cell, with a lattice constant of 10.604
$\mathrm{\AA}$,\citep{Y2O3} or a 40 atom primitive cell. 



Specifically, the 32 yttrium atoms in the cubic unit cell are
on the Wyckoff 8a
$(\frac{1}{4},\frac{1}{4},\frac{1}{4})$ 
and 24d $(u,0,\frac{1}{4})$ sites, with oxygen atoms at 48e
$(x,y,z)$. Previously obtained values of lattice parameter ($a$) $u, x, y, z$ are given in
table \ref{xyzu}.\citep{160890,181825,181871}

\begin{table}[H]
\begin{center}
\begin{tabular}{| c | c| c c c c | }
\hline
{Method} & $a$ & $x$ & $y$ & $z$ & $u$ \\ \hline
{X-Ray Diffraction \citep{160890}} & 10.6 & 0.3895(4) & 0.1509(3) & 0.3820(4) & 0.03203(6)\\
{Empirical potential \citep{181825}} & 10.6 &0.3906 & 0.1513 & 0.3797 & 0.0325 \\ 
{Density Functional Theory \citep{181871}} &10.46 & 0.3903 & 0.1506 & 0.3805 & 0.0283 \\ 
This work & 10.7& 0.3910 &0.1516 & 0.3798  &0.0326  \\ \hline
{Approximated} & &$\frac{3}{8}$ & $\frac{1}{8}$ & $\frac{3}{8}$ & 0 \\ \hline
\end{tabular}
\end{center}
\caption{\label{xyzu}Comparison of structural parameters for yttrium oxide obtained through various methods\citep{CDS}. Rounding to the nearest $\frac{1}{8}$ the parameters can be approximated.}
\end{table}

\begin{figure}[H]    
\begin{center}
\includegraphics[width=60mm]{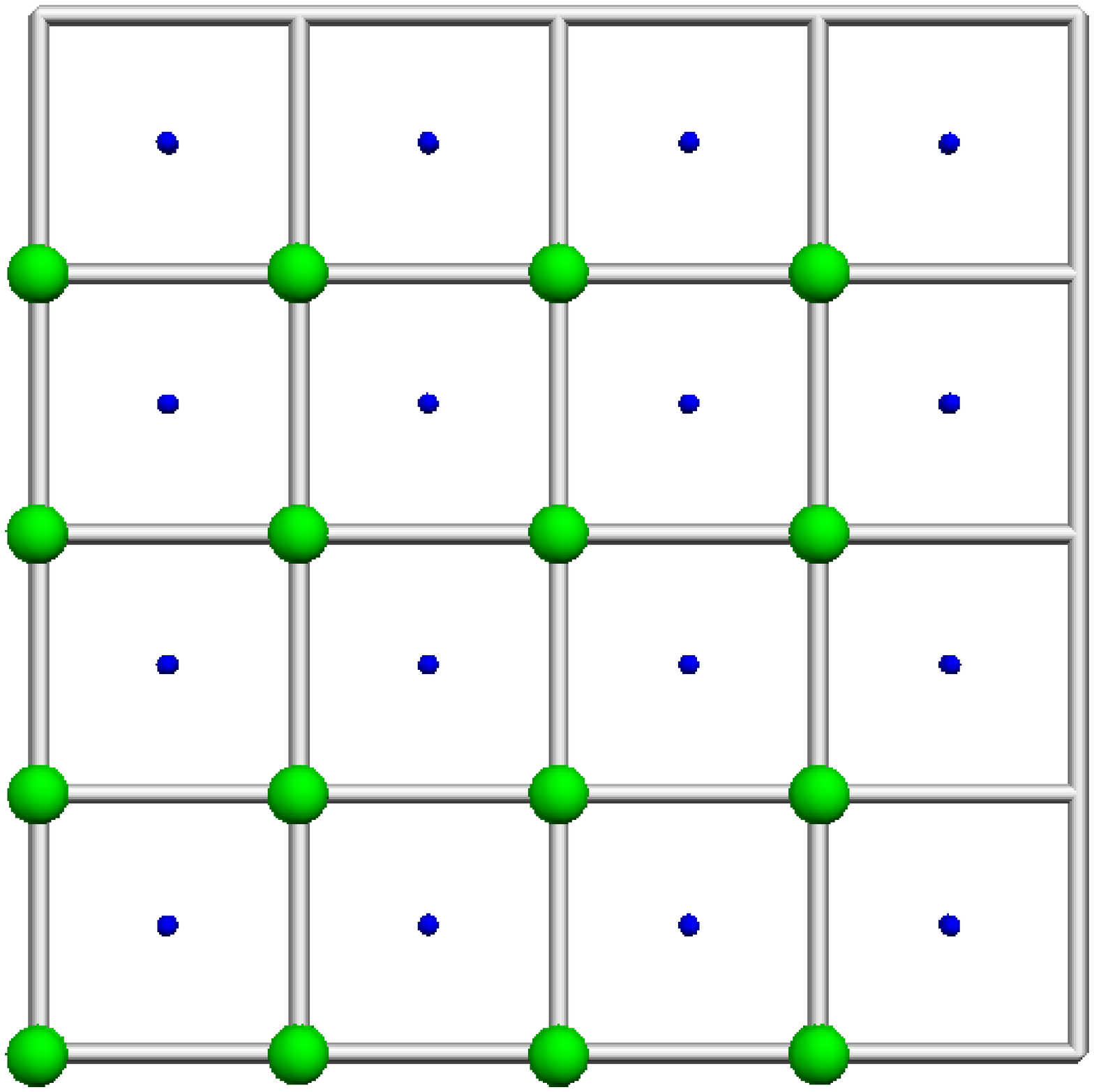}
\hspace{5mm}
\includegraphics[width=60mm]{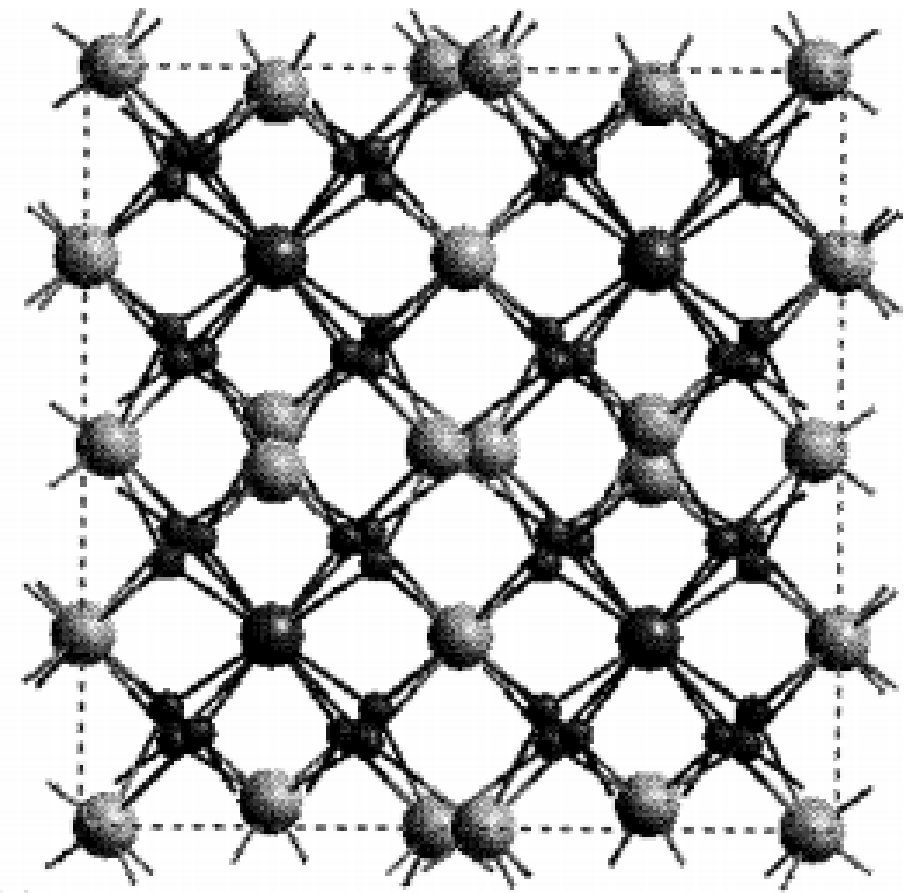}
\end{center}
\caption{\label{fig:Y2O3}The yttrium oxide unit cell can be thought of as atoms in the centres and vertices of the cubes forming a 4x4x4 grid. The idealised (left) and the actual (right) structure\citep{Y2O3} are visualised above. In both diagrams the larger atoms represent yttrium (green) and the smaller oxygen (blue).}
\end{figure}

Note that the structural parameters are quite close to integer
multiples of $\frac{1}{8}$. Rounding these to the nearest such values
approximates the yttrium and oxygen atoms to be located on vertices
and centres, respectively, of cubes comprising a 4x4x4 grid. From this
viewpoint the oxygen atoms occupy two thirds of the possible 64 sites,
while the yttrium atoms occupy half (Fig. \ref{fig:Y2O3}) 

\subsection{Interface Characterisation}\label{sec:IntChar}
In order to account for the observed radiation resistant properties of
yttria-based ODS steels, the microstructure of the oxide dispersions
in bulk iron must be considered. The dispersion of the
$\mathrm{Y_2O_3}$ nanoparticles throughout the steel has been observed
to be largely homogeneous,\citep{microstruc} with a typical density of
about one nanoparticle per $10^4\ \mathrm{nm}^3$\citep{castro}.
Generally, the nanoparticles are spherical with a diameter on the
order of ten nanometres, but larger nanoparticles tend to form
pronounced faceted surfaces with no evidence of large interfacial strain\citep{decanano}. 



Orientation correlations of the nanoparticle with the
steel matrix have been observed using High Resolution Transition
Electron Microscopy (HRTEM)\citep{decanano,hsiung,feal}, and for the
EUROFER97-based ODS steel the dominant
orientation correlation is `$[110]_{Y_2O_3}||[111]_{Fe}$ and
$(1\bar{1}\bar{1}) _{Y_2O_3}||(1\bar{1}0)_{Fe}$', where the aligned
directions and Miller indices of the interface are specified.



Detailed measurement of energetics of the interface and its
interactions are needed to show whether the nanoparticles act as
attractive sinks for point defects.  Here we use DFT to investigate
this for two orientation correlations: the observed EUROFER97 correlation
specified above and a simpler `$[010]_{Y_2O_3}||[010]_{Fe}$ and $(100)
_{Y_2O_3}||(100)_{Fe}$' which also involves low strain mismatch.

\subsubsection{Structure of  Simple Interface $\mathbf{[010]_{Y_2O_3}||[010]_{Fe}}$ and $\mathbf{(100) _{Y_2O_3}||(100)_{Fe}}$}

This interface is constructed by simply placing the two cubic
structures side by side and joining them at a cube face.
The
ratio between the yttrium oxide and iron lattice constants is 3.7. This
suggests a supercell with the periodicity
of the $\mathrm{Y_2O_3}$ cubic unit cell matches $4\times 4$ bcc iron cells. The
strain in each structure from
this fit is 4.0\%.
The orientation is uniquely defined due to the cubic 
symmetry, but the truncation of the  $\mathrm{Y_2O_3}$ slab may be either at 
O or Y atoms.
Thus, the two cases
of the iron bonding to the oxygen or yttrium must both be
investigated, let these interfaces be referred to as (100)Fe-O and
(100)Fe-Y respectively. A schematic of how these two setups might be
constructed is shown below. Note the continuation of the bcc structure
into the interface.

\begin{figure}[H]    
\centering
\begin{tabular}{c@{\hspace{5mm}}|l|@{\hspace{5mm}}c}
{\bf(100)Fe-O}
& \multicolumn{1}{c|@{\hspace{5mm}}}{\bf Key:}
& {\bf(100)Fe-Y} \\
\multirow{6}{*}{\includegraphics[width=60mm]{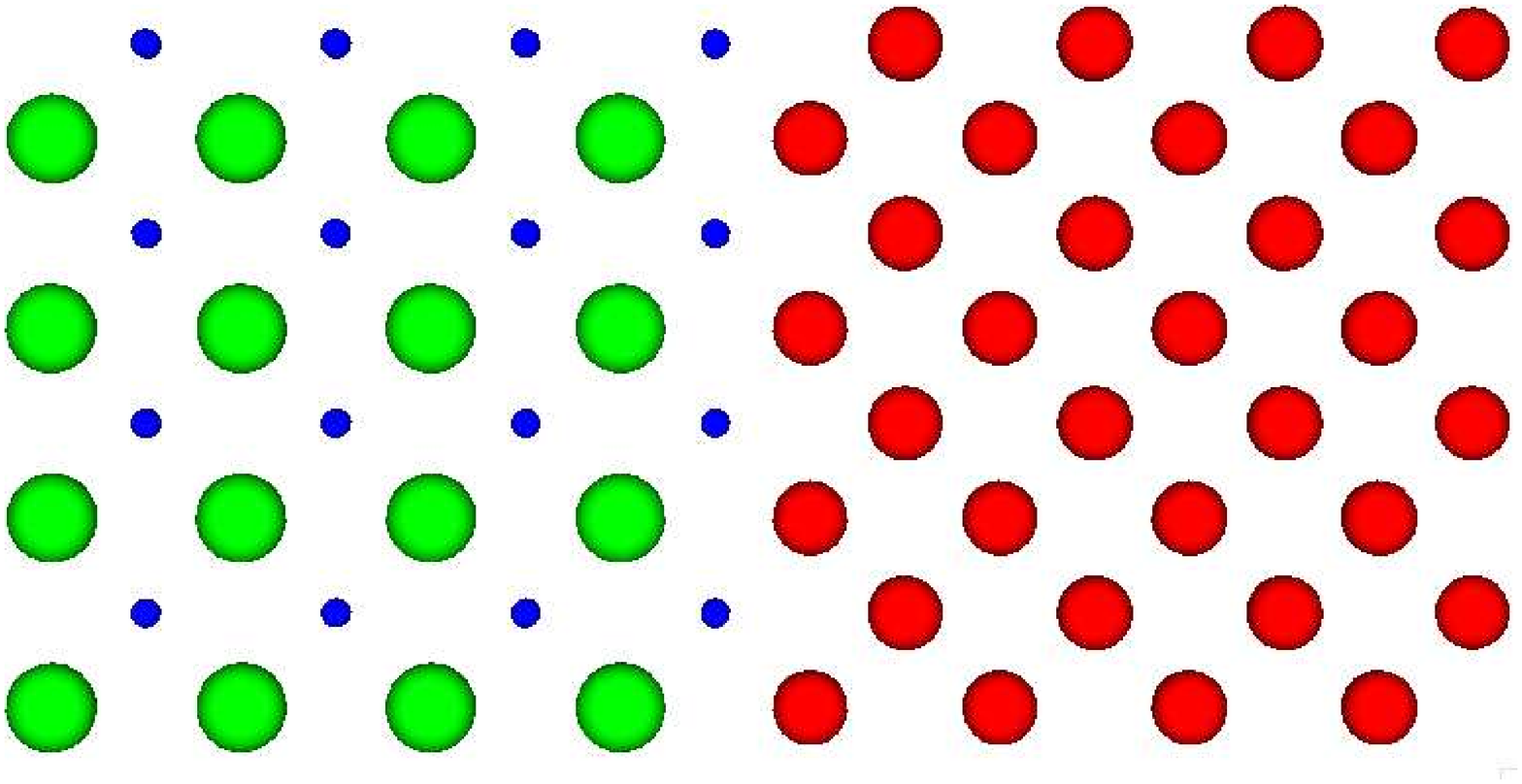}} 
&
& \multirow{6}{*}{\includegraphics[width=60mm]{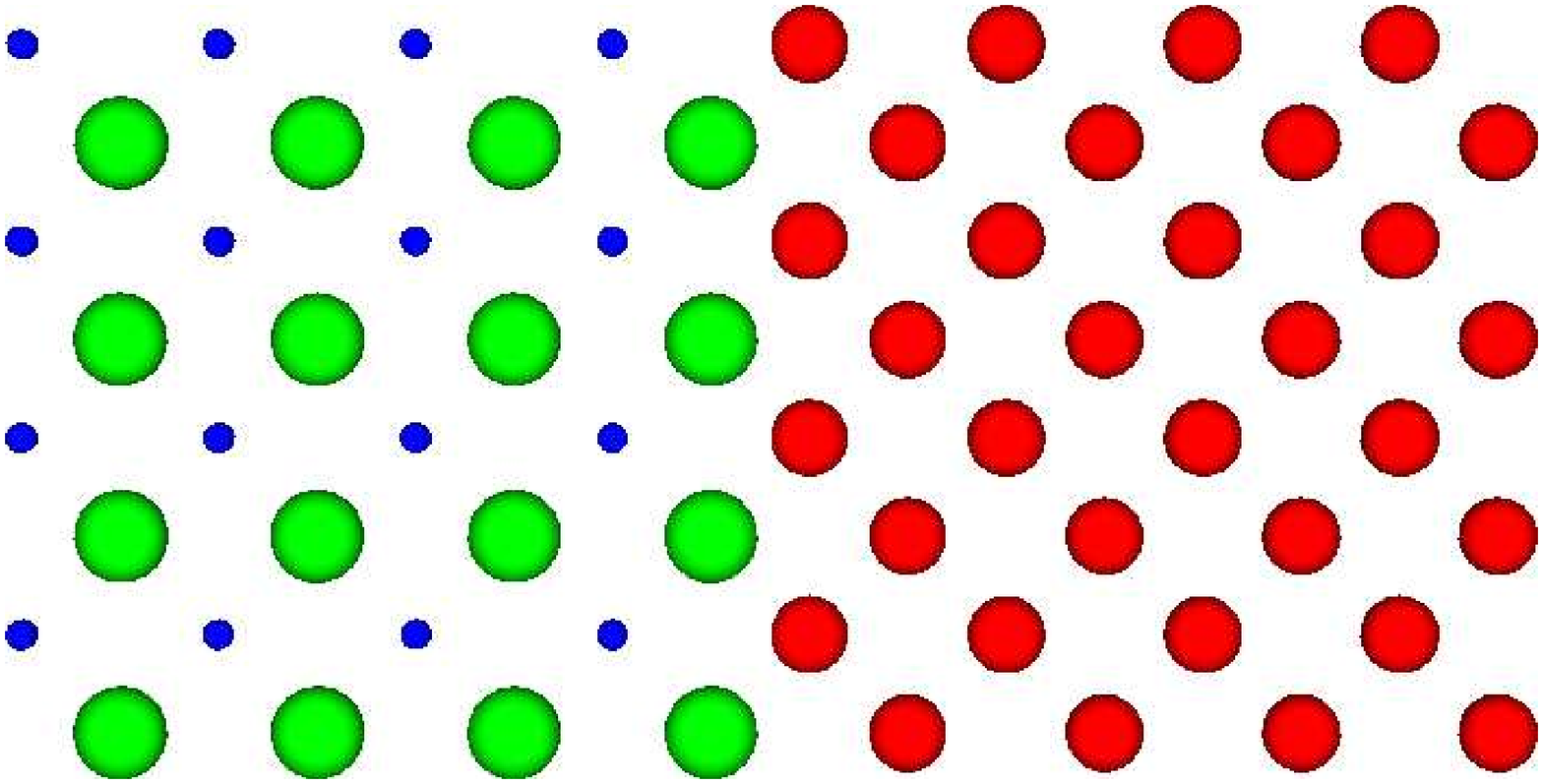}}\\
& \includegraphics[width=3mm]{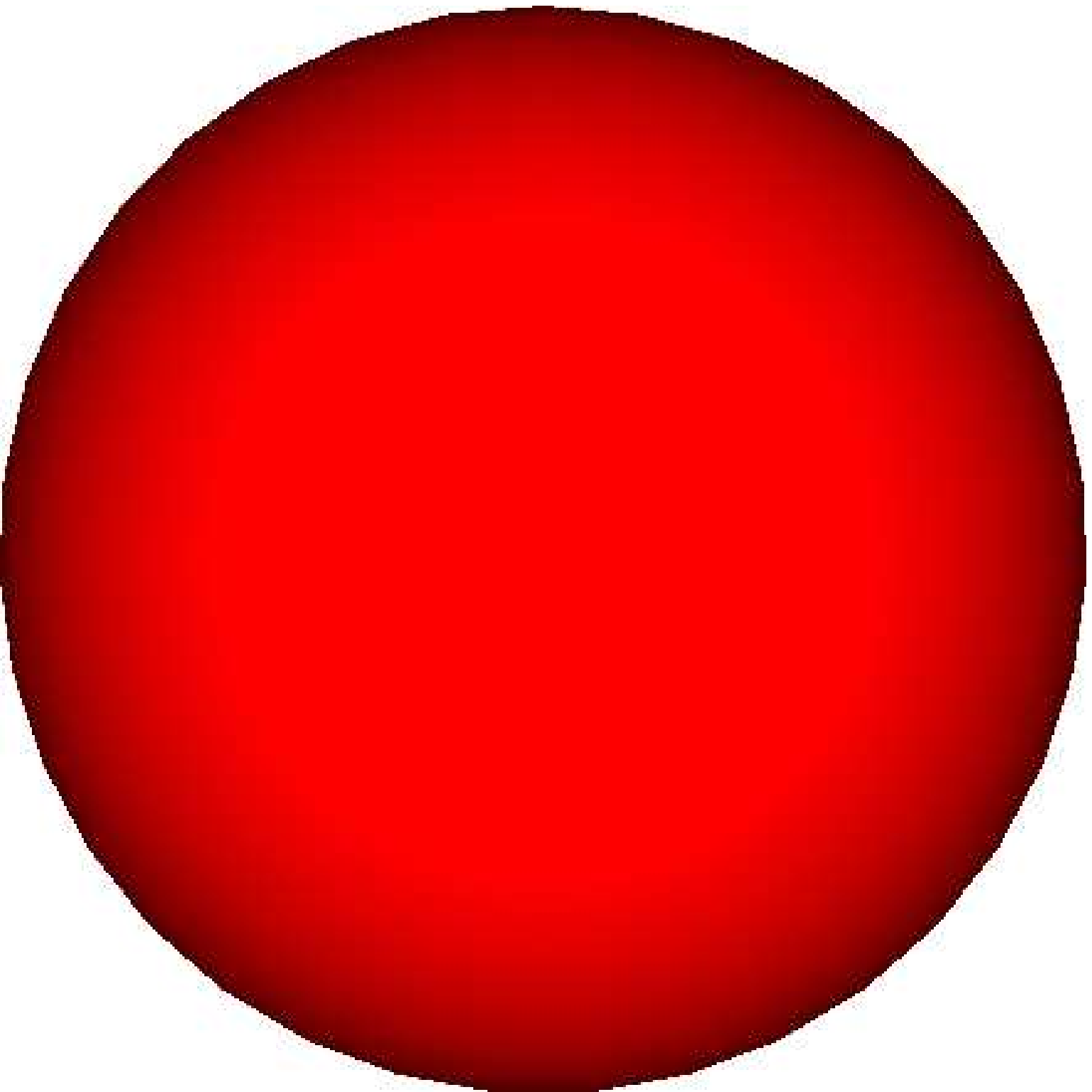} Iron  & \\
& \includegraphics[width=3mm]{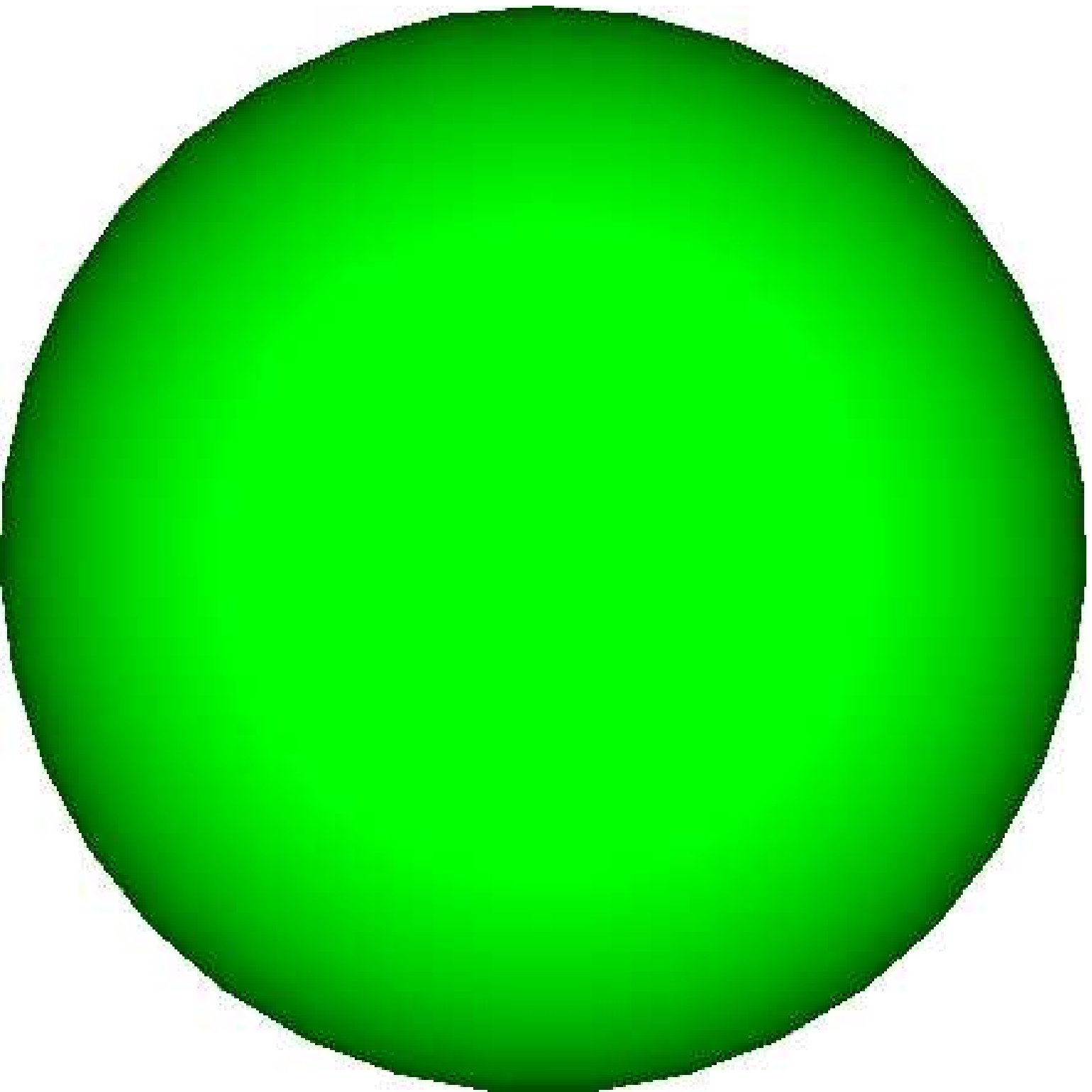} Yttrium & \\
& \includegraphics[width=3mm]{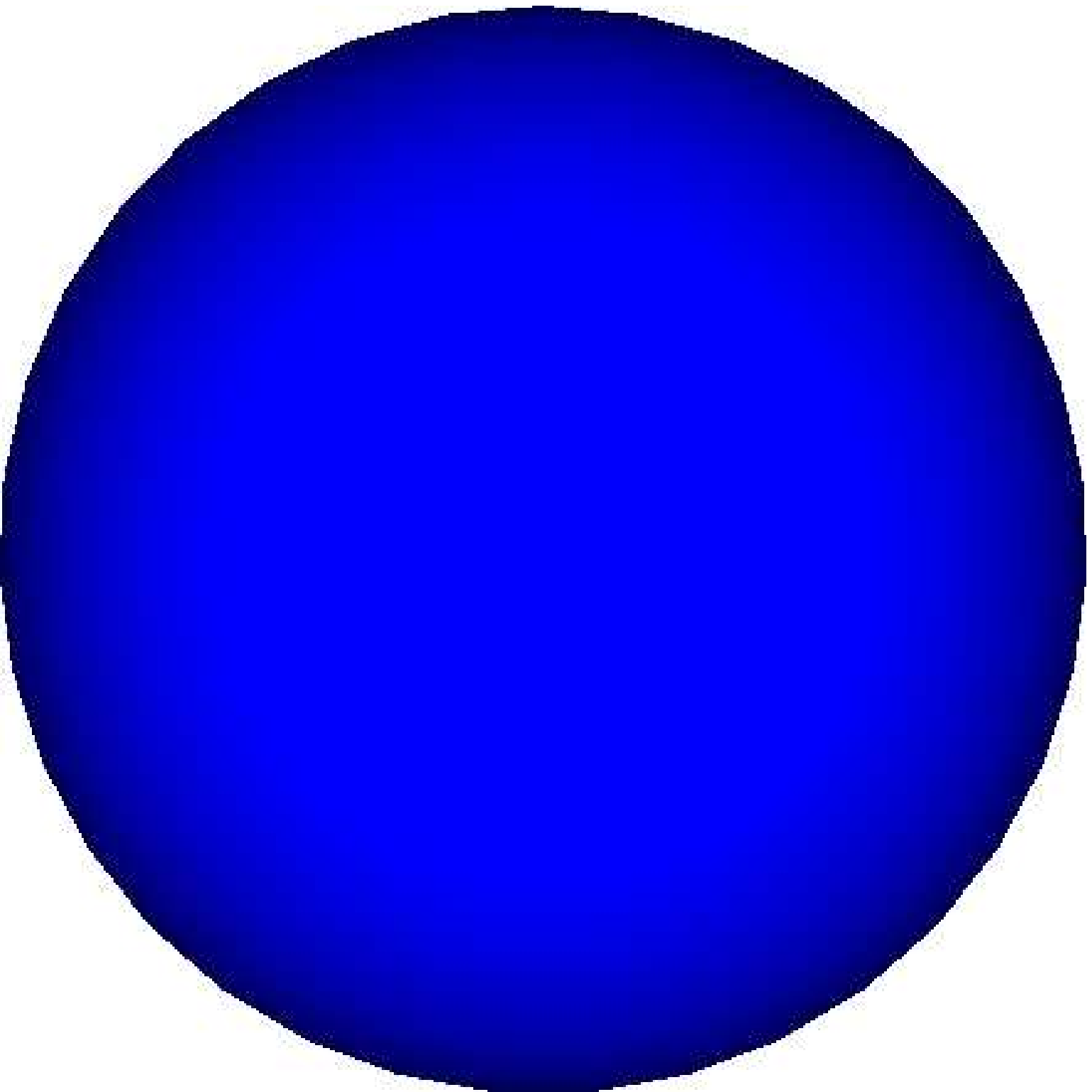} Oxygen  & \\
& & \\
& & \\
\end{tabular}
\caption{Both iron-oxygen and iron-yttrium interfaces are possible.}
\end{figure}

\subsubsection{Structure of  Klimiankou Interface: $\mathbf{[110]_{Y_2O_3}||[111]_{Fe}}$ and $\mathbf{(1\bar{1}\bar{1}) _{Y_2O_3}||(1\bar{1}0)_{Fe}}$}

The orientation of the interface observed in EUROFER97\citep{decanano}
is slightly more complicated as the two planes are of different types.
The interface model is best visualised by cutting the cubic unit
cells along the relevant plane and shifting by a lattice vector,
forming parallelepipeds each with two faces of the required type. If an
arrow is drawn along the correlated directions on the respective
faces, the interface is oriented by making both these arrows
parallel. Again, due to symmetry, all orientations
`${<}111{>}_{Y_2O_3}||{<}110{>}_{Fe}$ and $\{110\}
_{Y_2O_3}||\{111\}_{Fe}$' type are accounted for by this specific
orientation correlation.

\begin{figure}[H]    
\begin{center}
\includegraphics[width=60mm]{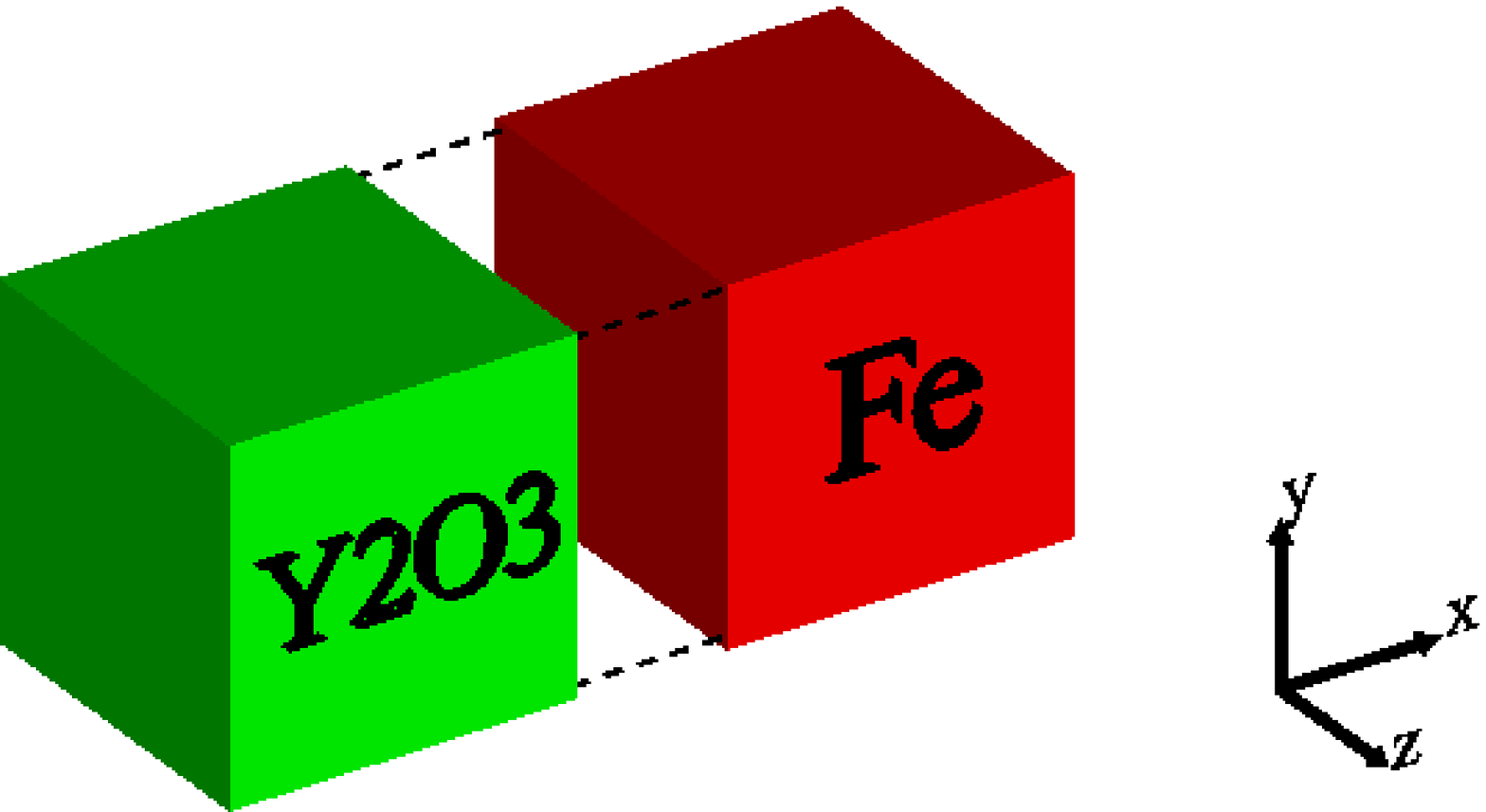}
\includegraphics[width=60mm]{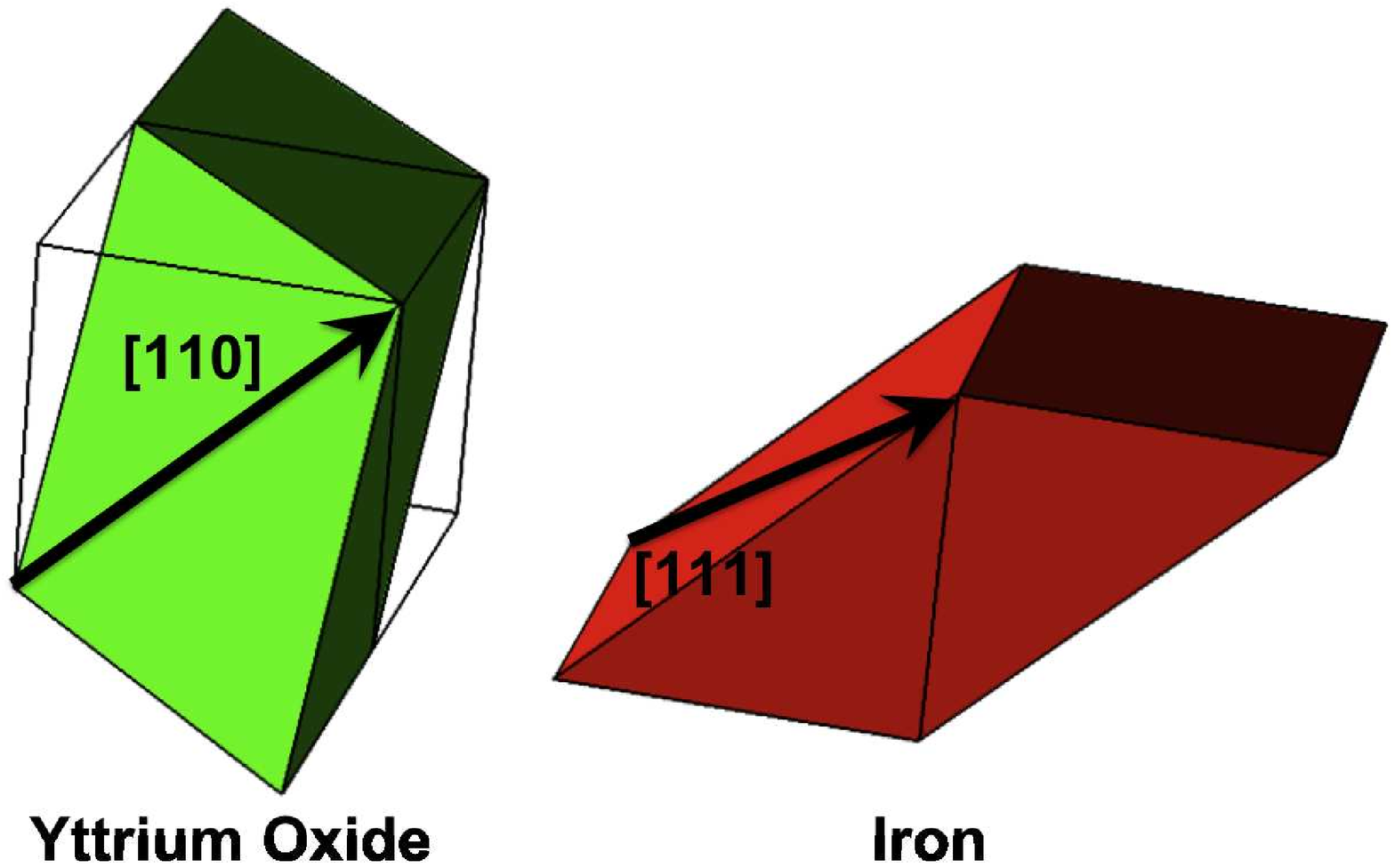}
\end{center}
\caption{\label{fig:Klim3D}The (100) interface is obtained by placing
  the cubic structures side by side and joining two faces. For the
  Klimiankou geometry\citep{decanano} the yttrium oxide and iron unit cells are cut
  along the $(1\bar{1}\bar{1})$ and $|(1\bar{1}0)_{Fe}$ respectively
  to form parallelepipeds. The interface is formed by joining the
  parallelepipeds at these faces and making the $[110]_{Y_2O_3}$ and
  $[111]_{Fe}$ directions parallel.}
\end{figure}

\begin{figure}[H]
\begin{center}
\begin{tabular}{c@{\hspace{5mm}}|@{\hspace{5mm}}c@{\hspace{5mm}}|@{\hspace{5mm}}c}
{\bf Fe-Y-O} & {\bf Fe-O-Y} & {\bf Fe-O-O} \\[3mm]
\includegraphics[width=45mm]{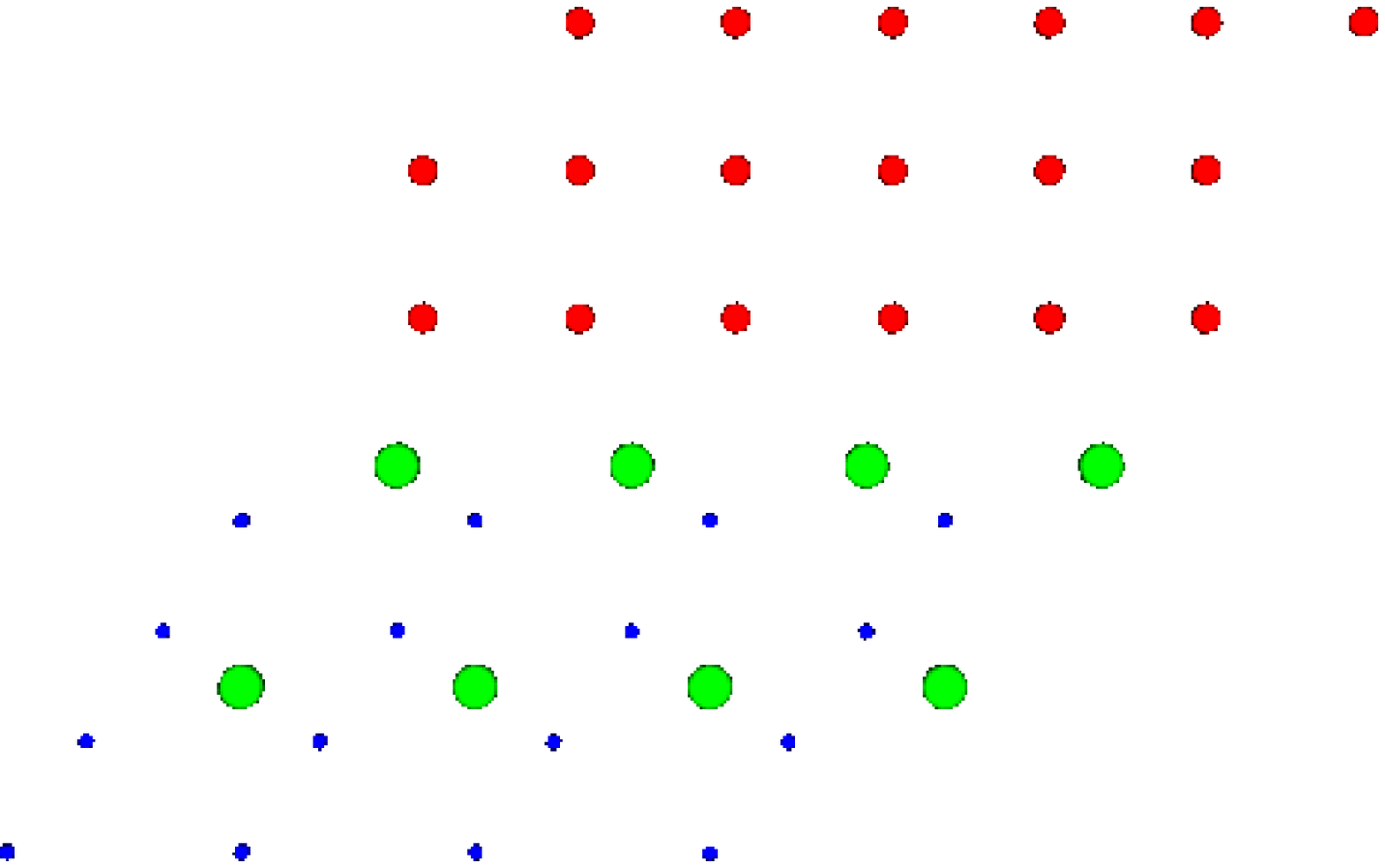}
& \includegraphics[width=45mm]{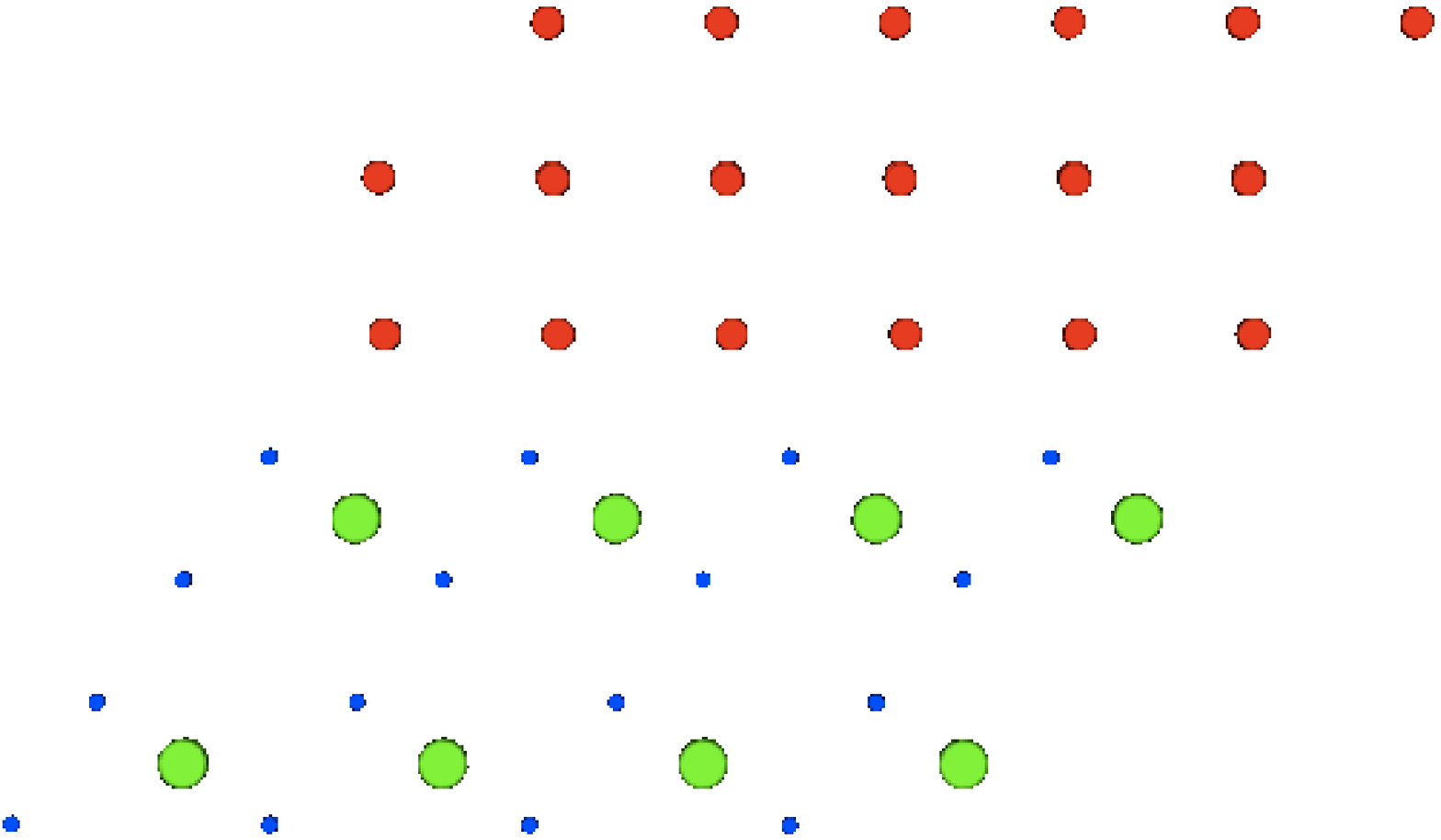}
& \includegraphics[width=45mm]{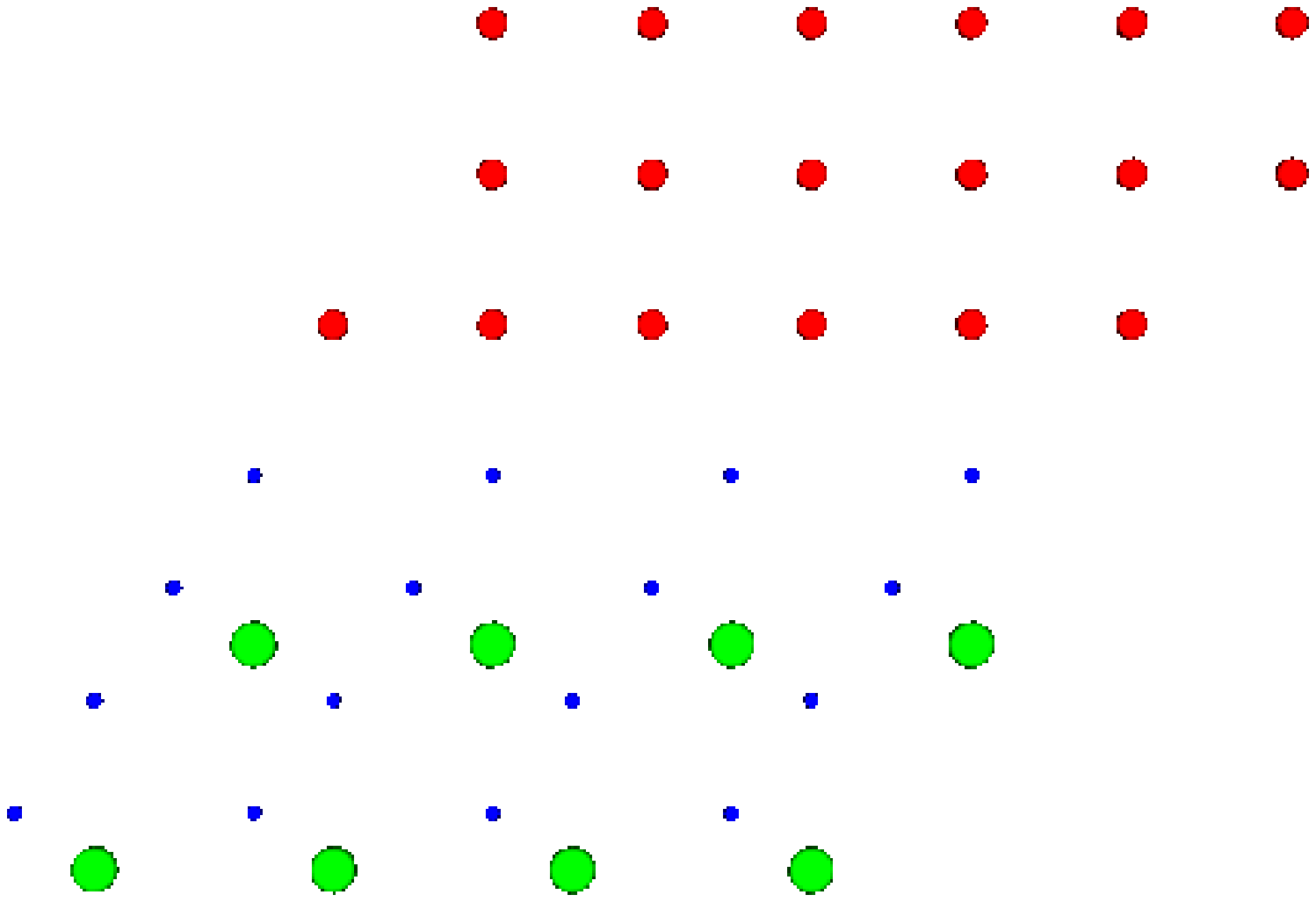} \\
\end{tabular}
\end{center}
\caption{Three distinct interface types are possible and are depicted here viewed along the [01$\bar{1}$] axis.\label{KlimPlane} }
\end{figure}

There are two distinct
planes of the $(1\bar{1}\bar{1})$ type in  $\mathrm{Y_2O_3}$, one containing 
16 yttrium atoms and the other with 12 oxygens. 
They can be thought of as
atoms located on the vertices of a $4\times4$ parallelogram grid, with
some vertices empty in the case of oxygen.
Consequently, iron-yttrium and iron-oxygen interfaces are both
possible.  Furthermore, there are two varieties of the iron-oxygen
interface: The next layer into the oxide could be either a yttrium or
another oxide layer. We label the candidate interfaces KlimFe-Y-O,
KlimFe-O-Y and KlimFe-O-O, as depicted in figure
\ref{KlimPlane}.

The ratio between lengths in the correlated directions is
$\frac{\sqrt{2}\times10.60}{\sqrt{3}\times2.87} = 3.02$. Thus a strain
of only 0.4\%, is required with three iron parallelepipeds joined
corner-to-corner along the arrows depicted in figure
\ref{fig:Klim3D}. 
In the
perpendicular direction the misfit can
be accommodated by a shear in the iron lattice resulting in an increase of
the angle between the $[111]_{Fe}$ and $[001]_{Fe}$ directions from
$55^\circ$ to $60^\circ$.

\begin{figure}[ht]
\begin{center}
\begin{tabular}{c|c|c}
Yttrium & Oxygen & Iron \\[3mm]
\includegraphics[height=32mm]{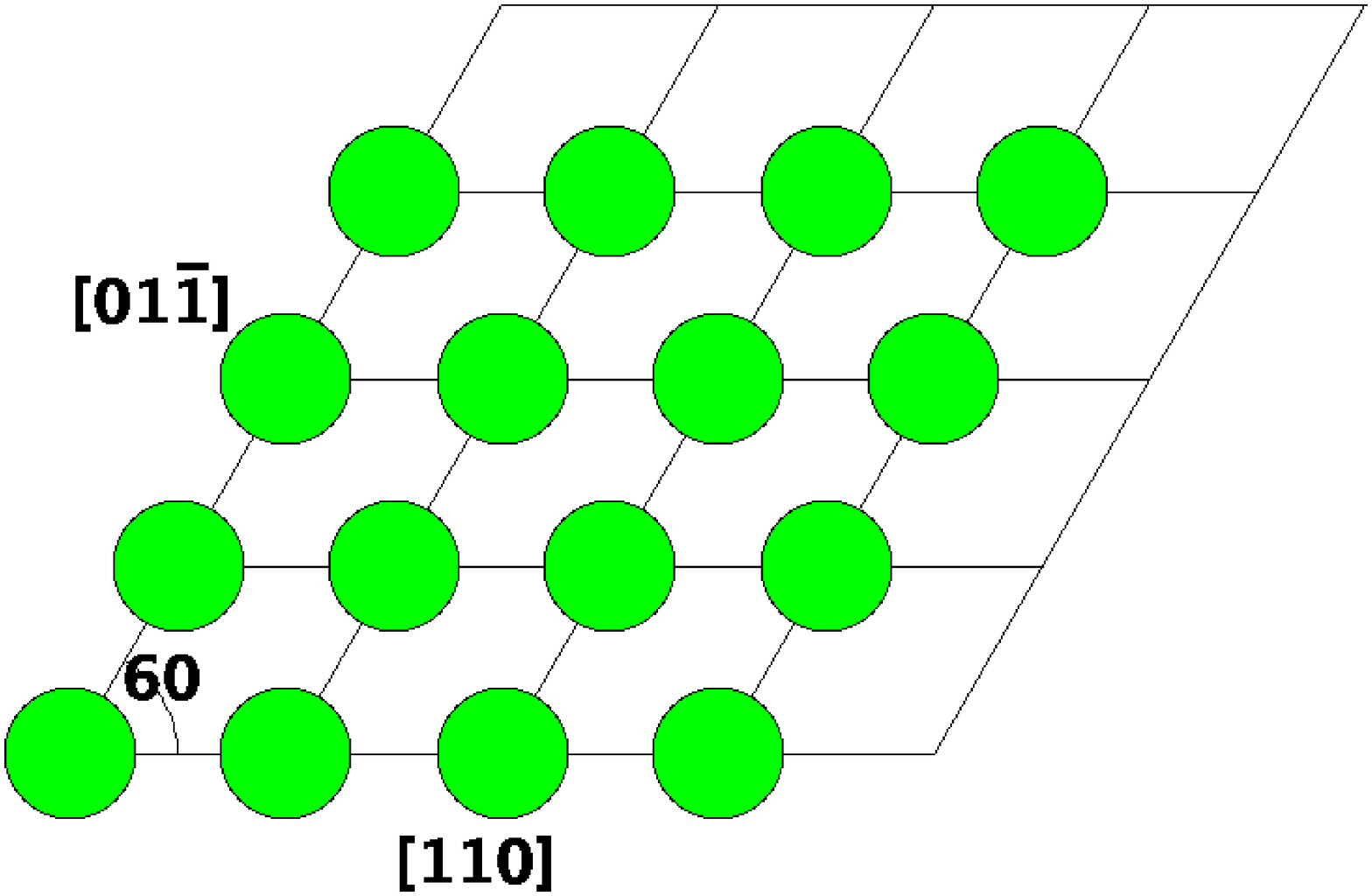}
& \includegraphics[height=32mm]{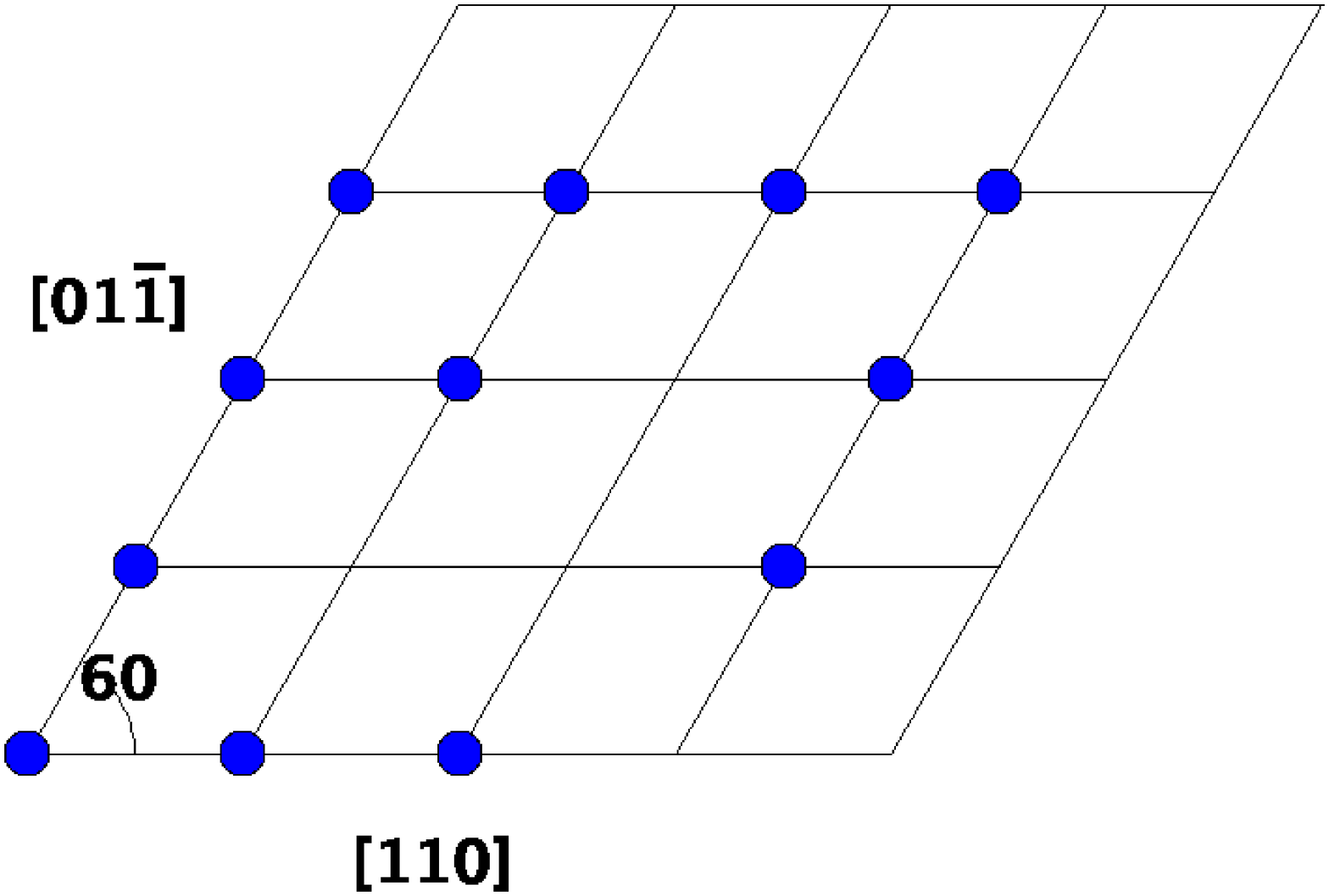}
& \includegraphics[height=30mm]{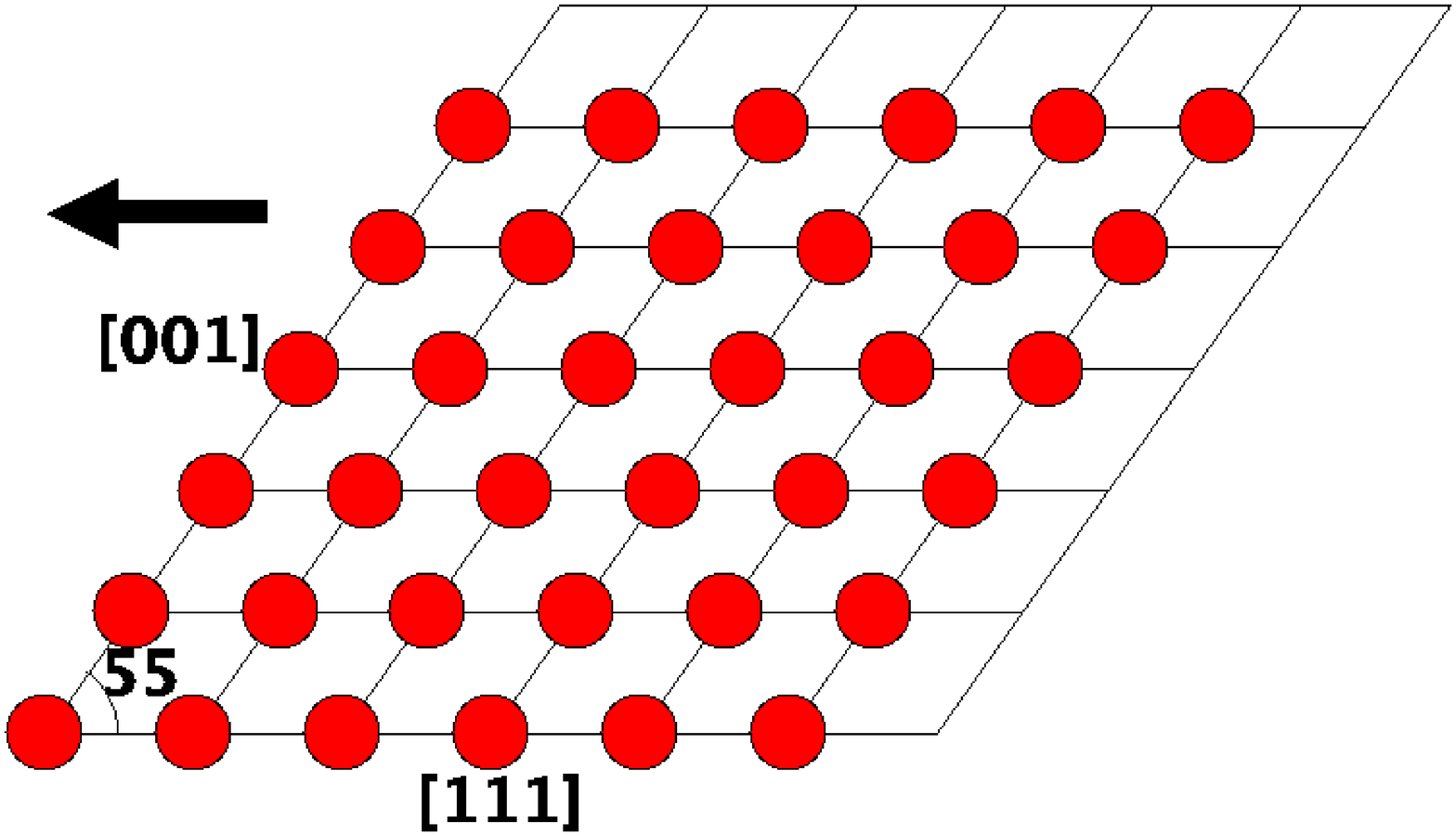} \\
\end{tabular}
\end{center}
\caption{Schematic of the element planes parallel to the Klimiankou
  interface as vertices of parallelogram grids. Note the shear
  necessary in the iron plane for the it to be congruent with yttrium
  oxide.  The interface is a (110) plane in Fe and a (21$\overline{1}$)
  plane in yttria.}
\end{figure}

\section{Computational Details}

To calculate the interfacial energies, we use the Density Functional
Theory as implemented in the Vienna Ab Initio Simulation
Package (VASP)\citep{VASP} using the projector augmented-wave (PAW)
method\citep{paw} with eight(Fe), six(O) and eleven(Y) valence electrons
and the spin polarised PBE exchange-correlation
functional\citep{PBE} with a 520eV cutoff energy for the plane wave basis set.
The interface
supercells contained over 200 atoms, and the largest calculations used
$1\times3\times3$ k-point Monkhorst-Pack mesh (with the lower
number of k-points in the longer direction).
All atoms were fully relaxed from the ``ideal'' geometries described above.
These settings are known from previous work to be reliable\citep{Fu,FeNiCr,Olsson,Zhu}.

The stability of the interface was determined by the
work of adhesion for stoichiometric setups. The effect of creating iron
vacancies and SIAs both at and near the interface was
investigated by comparing the relative energy change with that in pure
iron after ionic, but not volumetric, relaxation.
between consecutive steps was less than $10^{-4}$ eV.  The chemistry
of the binding was investigated from the density of states projected
on atoms near the interface.




To calculate the reference energy cost of forming vacancies and SIAs
in pure iron, we used a 256 atom supercell.  Lattice constant
(2.87\AA), magnetic moment (2.2$\mu_B$) vacancy ($E_F^{vac}$=2.16eV)
and (110)-SIA ($E_F^{SIA}$=4.02eV) energies in iron were consistent
with previous work\citep{Fu,FeNiCr,Olsson}.  The calculated lattice
parameter of Y$_\mathrm{2}$O$_\mathrm{3}$ is 10.7 $\mathrm{\AA}$, and
its structural parameters, $x, y, z, u$ are given in table \ref{xyzu}.
Our results are comparable to recent studies of Y and O
impurities in bcc FeCr\citep{Claisse,Jiang}. Y has a substitutional energy in
Fe of 2.01 eV and Oxygen 0.13eV relative to metal and $O_2$ gas state
respectively.


\section{Stoichiometry and Interfacial Energy}

Interfacial energy can be calculated when the yttrium to
oxygen ratio in the supercell is stoichiometric with yttrium oxide
i.e. 2:3. 
For example, if a supercell, with total      energy F, contains $N_{Fe}$ iron
particles and a total of $N_{Y_2O_3}$ yttrium and oxygen particles then
the associated interfacial energy would be:

\begin{equation}
\gamma = \frac{F-(N_{Fe}\ \mu_{Fe}^0+N_{Y_2O_3}\ \mu_{Y_2O_3}^0)}{A}
\end{equation}

Where \emph{A} is the total interfacial area in the supercell and
$\mu_{Fe}^0$=-8,212eV, $\mu_{Y_2O_3}^0=-9.164eV$ are the VASP chemical
potentials per atom for iron and Y$_\mathrm{2}$O$_\mathrm{3}$.  These absolute
values have no physical meaning, being the energy gained in forming the
perfect solid from a non-spin-polarised pseudoatom, but can be compared to the 
metallic Y (-6.26eV), molecular oxygen (-4.92eV) Y substitutional in Fe 
(-8.27eV) and O interstitial in Fe (-5.05eV) to show that Y and O will dissolve in Fe, but would prefer to form the oxide.



\section{Supercell Geometries}

\subsection{(100) Interfaces}
\begin{figure}[H]    
\centering
\begin{tabular}{c@{\hspace{5mm}}|l|@{\hspace{5mm}}c}
{\bf(100)Fe-O}
& \multicolumn{1}{c|@{\hspace{5mm}}}{\bf Key:}
& {\bf(100)Fe-Y} \\
\multirow{6}{*}{\includegraphics[width=40mm]{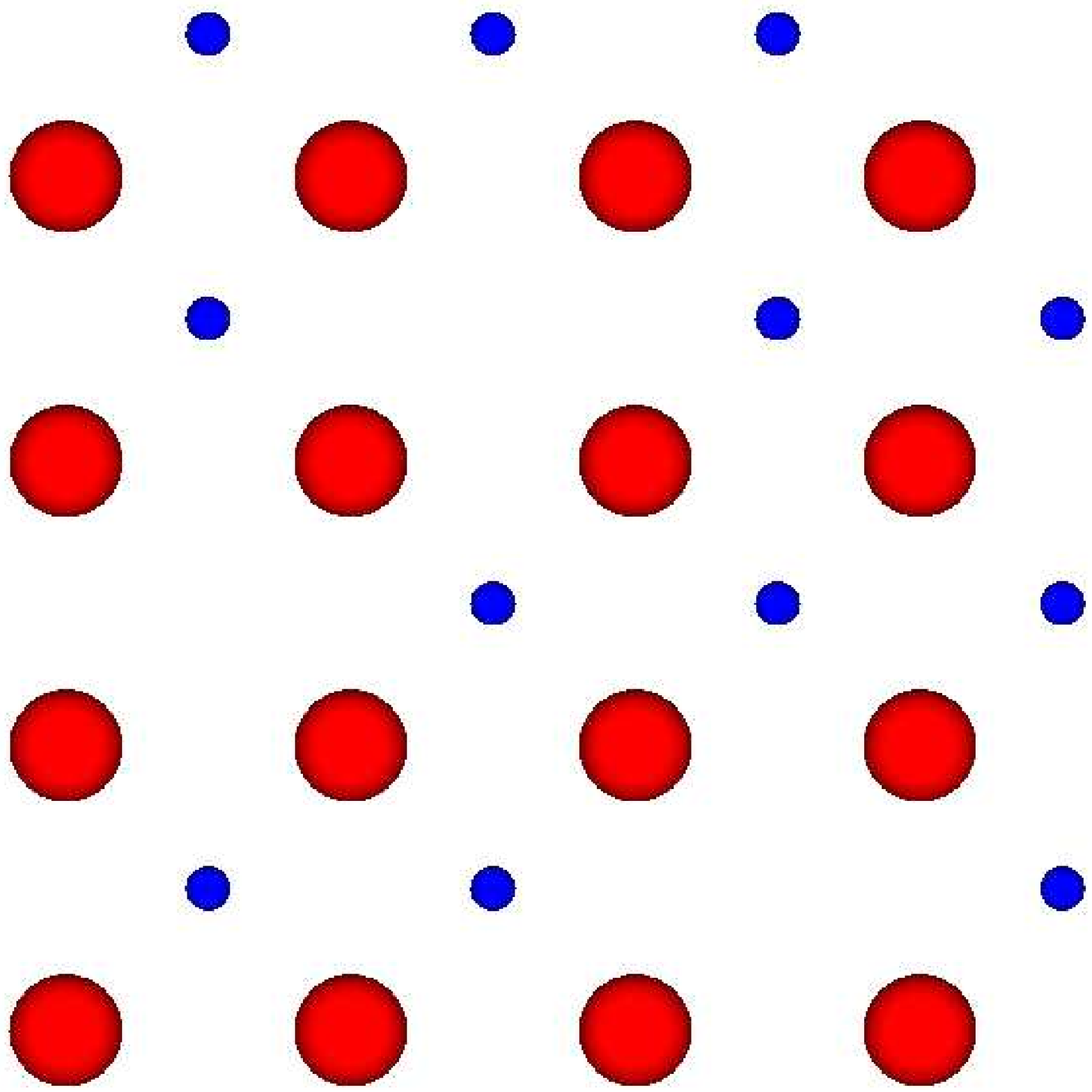}} 
&
& \multirow{6}{*}{\includegraphics[width=40mm]{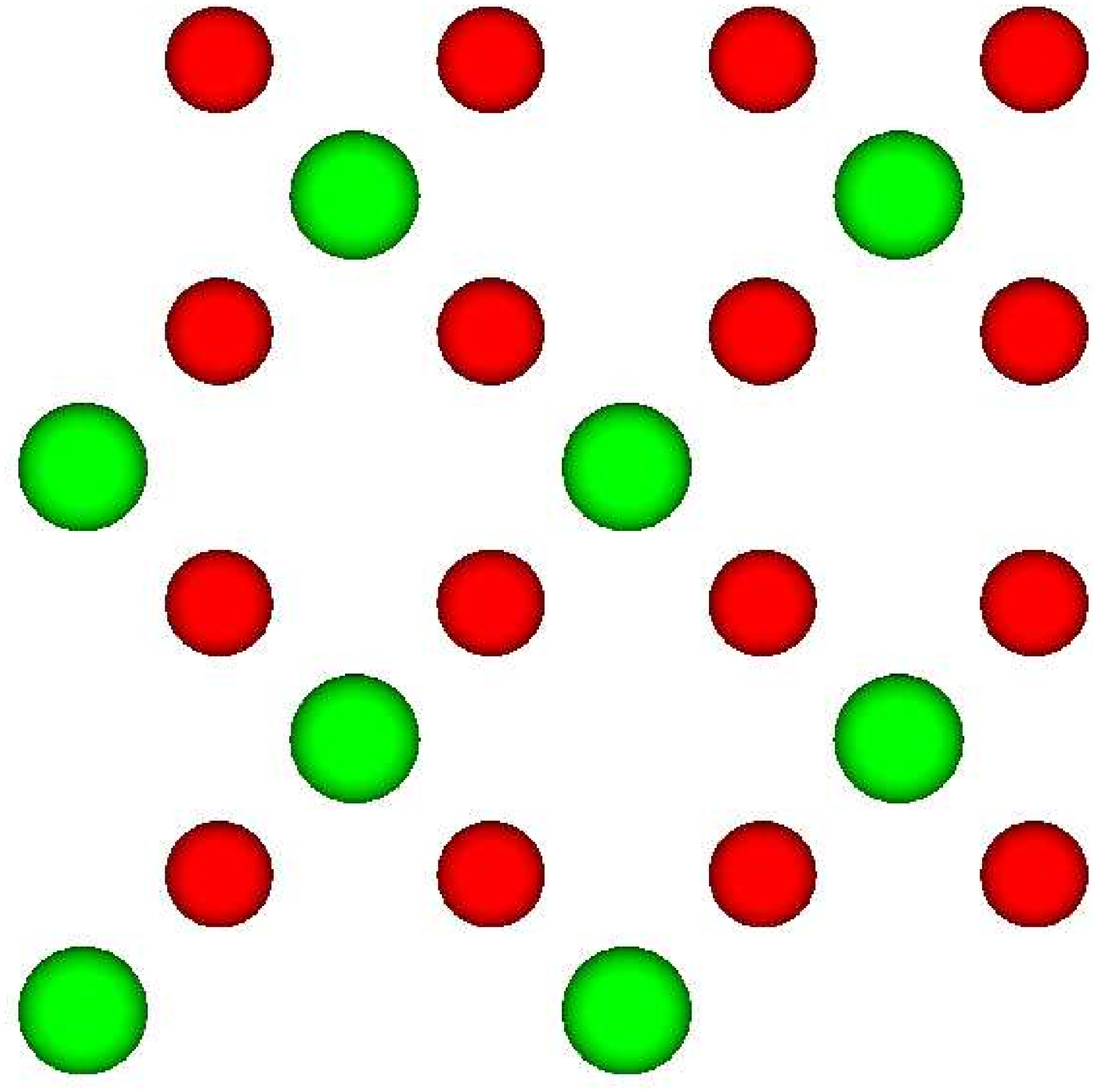}}\\
& \includegraphics[width=3mm]{Fe_Key.eps} Iron  & \\
& \includegraphics[width=3mm]{Y_Key.eps} Yttrium & \\
& \includegraphics[width=3mm]{O_Key.eps} Oxygen  & \\
& & \\
& & \\
& & \\
& & \\
\end{tabular}
\centering
\begin{tabular}{c@{\hspace{2mm}}|@{\hspace{2mm}}c@{\hspace{2mm}}|@{\hspace{2mm}}c}
{\bf(100)Fe-O}
& {\bf (100)Mixed}
& {\bf(100)Fe-Y} \\
 \includegraphics[height=28mm]{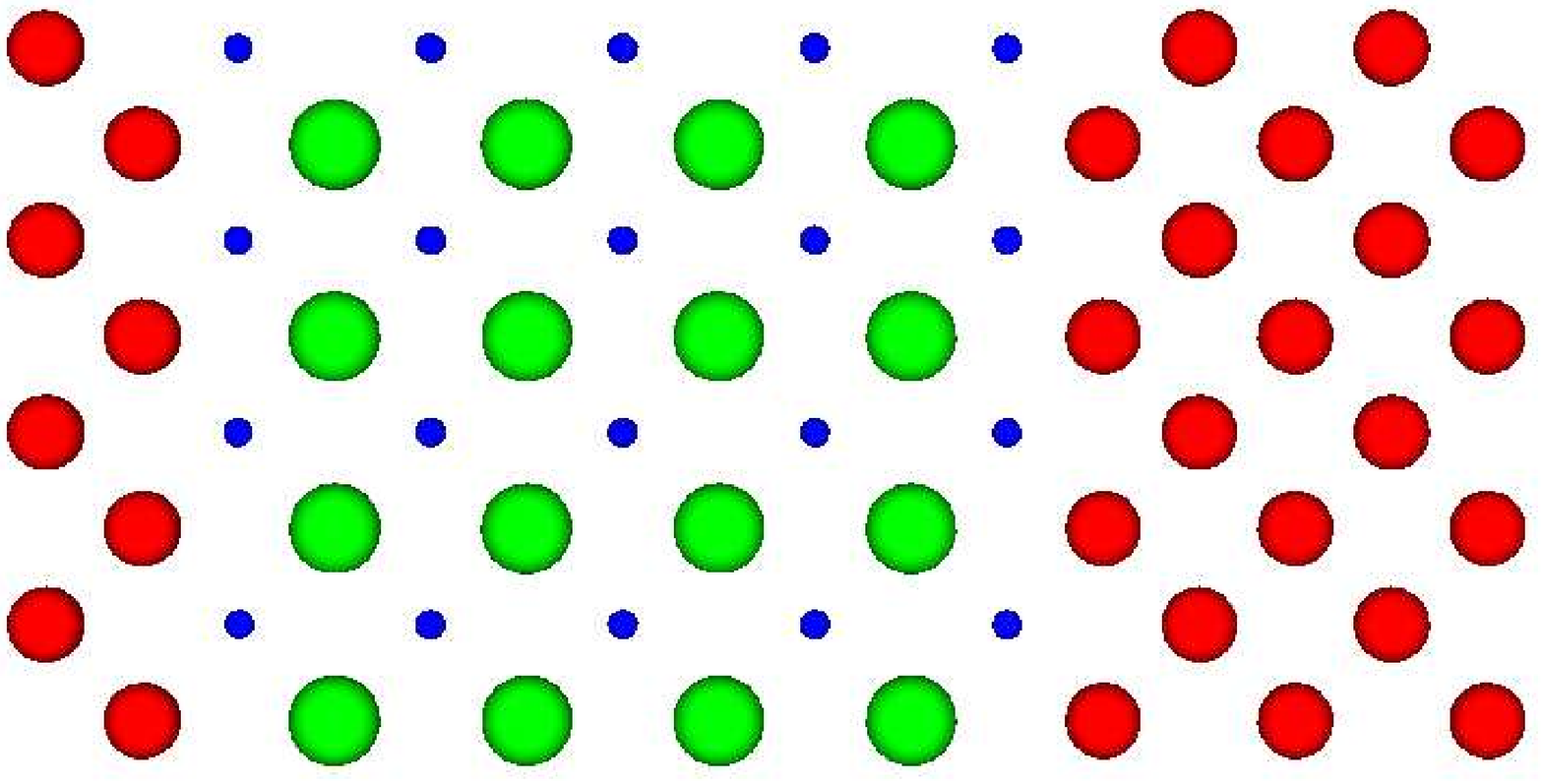} 
& \includegraphics[height=28mm]{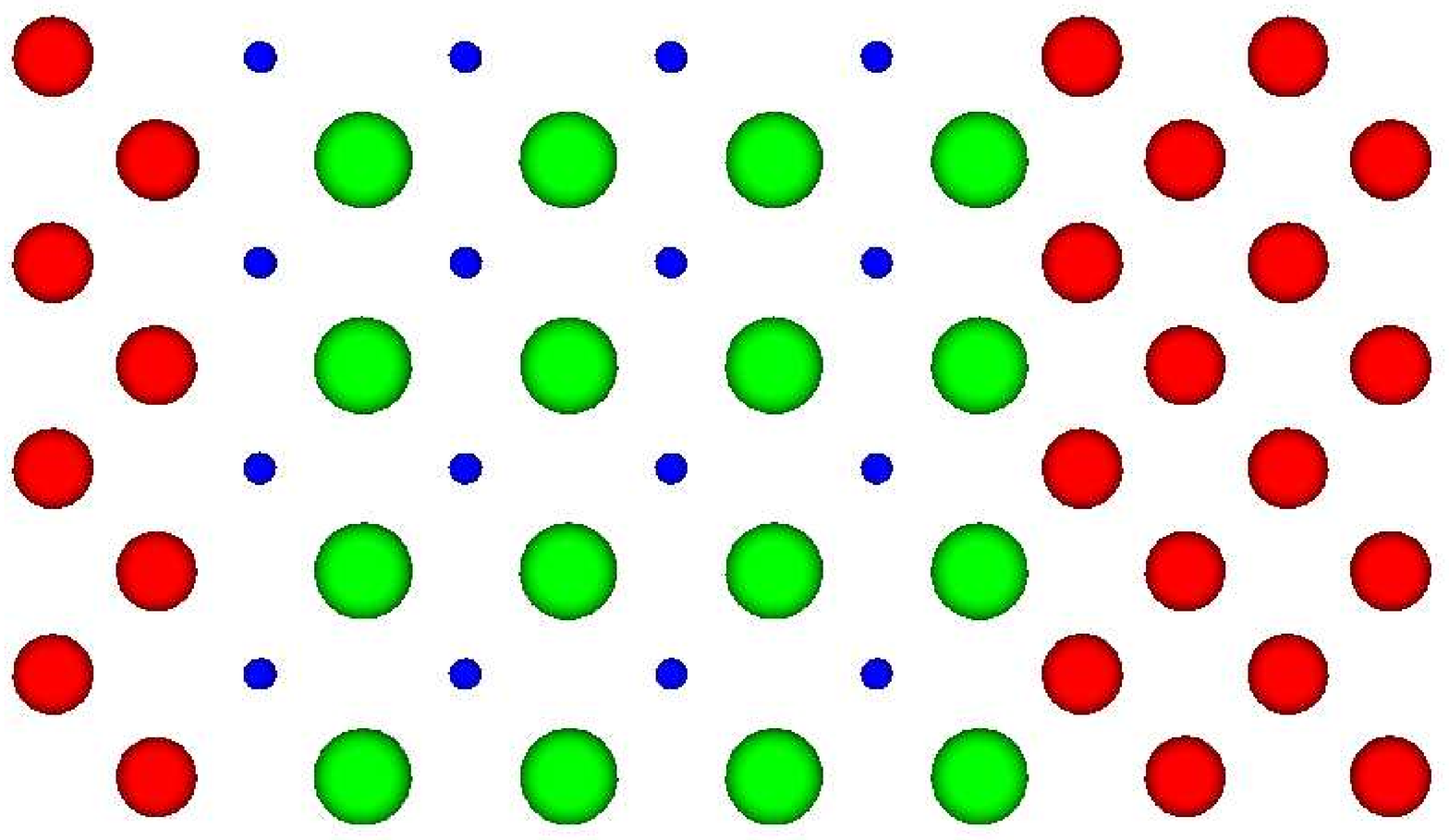} 
& \includegraphics[height=28mm]{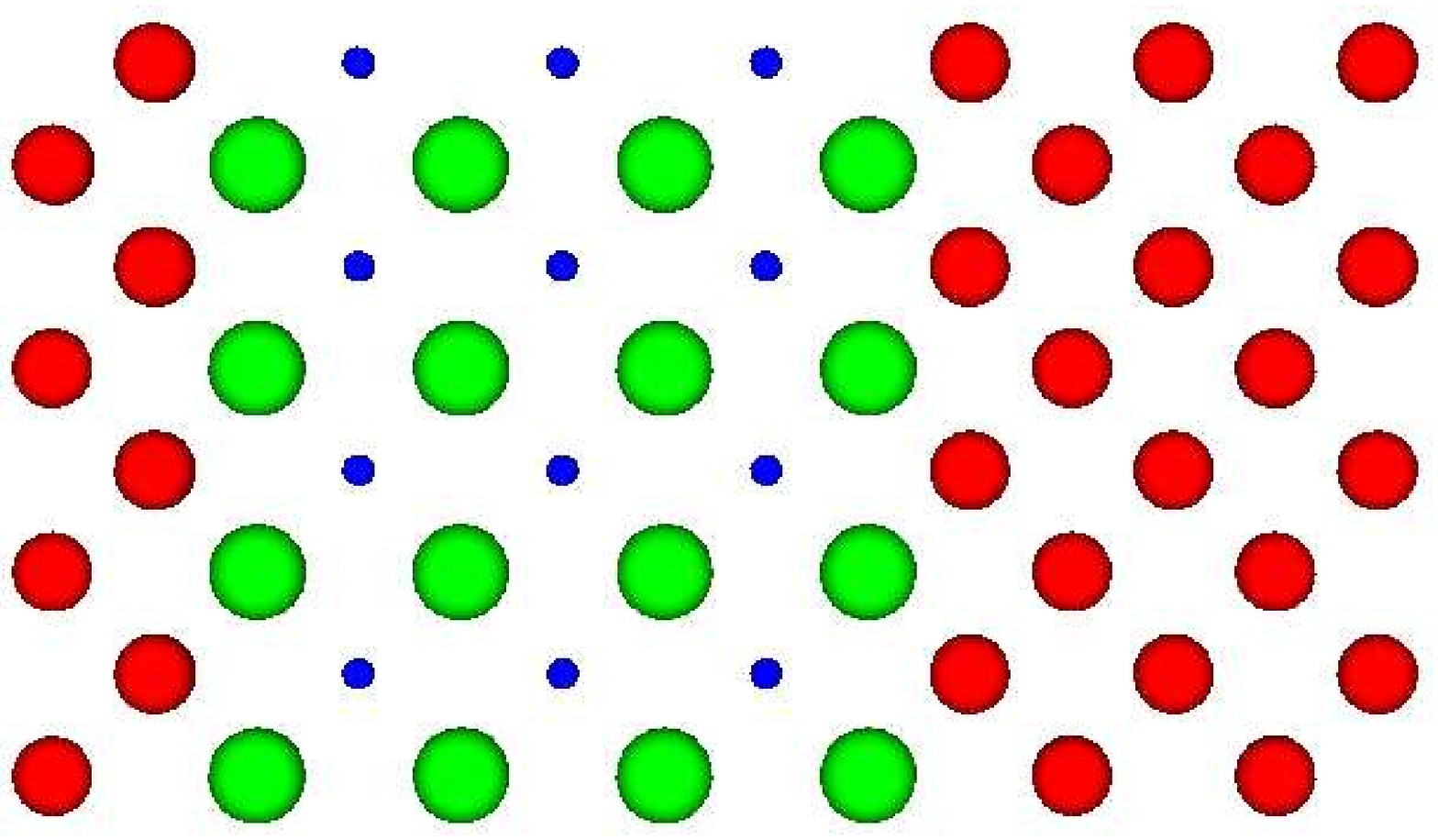}  \\
\end{tabular}
\caption{\label{fig:(100)Plane} (top) Layers above and below the iron-oxygen
  and iron-yttrium interfaces, viewed along the [100] axis. 
(bottom) Slice of supercell viewed
  along the [001] axis. Iron
  atoms have been placed at all sites in the interfacial layer. }
\end{figure}

Three supercells involving (100) interfaces were investigated: one
considering two (100)Fe-O interfaces, one considering two (100)Fe-Y
interfaces and the third considering one of each. The single-type
setups are denoted by the corresponding interface, the mixed
interface is called (100)Mixed. Only the mixed interface
has the $Y_2O_3$ stoichiometry needed for calculating the surface
energy. 
Periodic boundary conditions are used, and at
the interfaces the iron atoms were placed to
continue the bcc structure into the yttrium
oxide as shown in figure \ref{fig:(100)Plane}. 
The number of atoms in each supercell is given in
table \ref{100nos}.

\begin{table}[H]
\begin{center}
\begin{tabular}{| c |  c c c |  }
\hline
\multicolumn{1}{c}{} & \multicolumn{3}{|c|}{Number of atoms} \\ \cline{1-4}
{Setup} & Iron & Oxygen & Yttrium  \\ \hline
(100)Fe-O & 112 & 60& 32 \\
(100)Fe-Y  & 112 & 36 & 32 \\ 
(100)Mixed& 96& 48&32 \\ \hline
KlimFe-O-Y & 180 & 48 & 32\\
KlimFe-Y-O  & 144 & 48 & 48\\ 
KlimMixed & 180 & 48 & 32 \\ \hline
\end{tabular}
\end{center}
\caption{\label{100nos}The contents of
  the supercells.}
\end{table}

\subsection{Klimiankou Interfaces}
Two variations of the HRTEM interfaces\citep{decanano} have also been
investigated: a stoichiometric setup with two single-O layer interfaces
(denoted KlimFe-O-Y),
a non-stoichiometric setup with two KlimFe-Y-O. Various setups 
containing KlimFe-O-O interfaces were attempted, but 
underwent massive spontaneous reconstruction indicating instability. 
However, one stoichiometric setup (KlimMixed), also containing a
KlimFe-Y-O interface was relaxed enough to achieve acceptable
formation energy results. The lattice vectors
of the supercell were in the [010], [110] and [01$\bar{1}$]
directions. The number of atoms in each cell are given in
table \ref{100nos}.

\begin{figure}[H]
\begin{center}
\includegraphics[width=51mm]{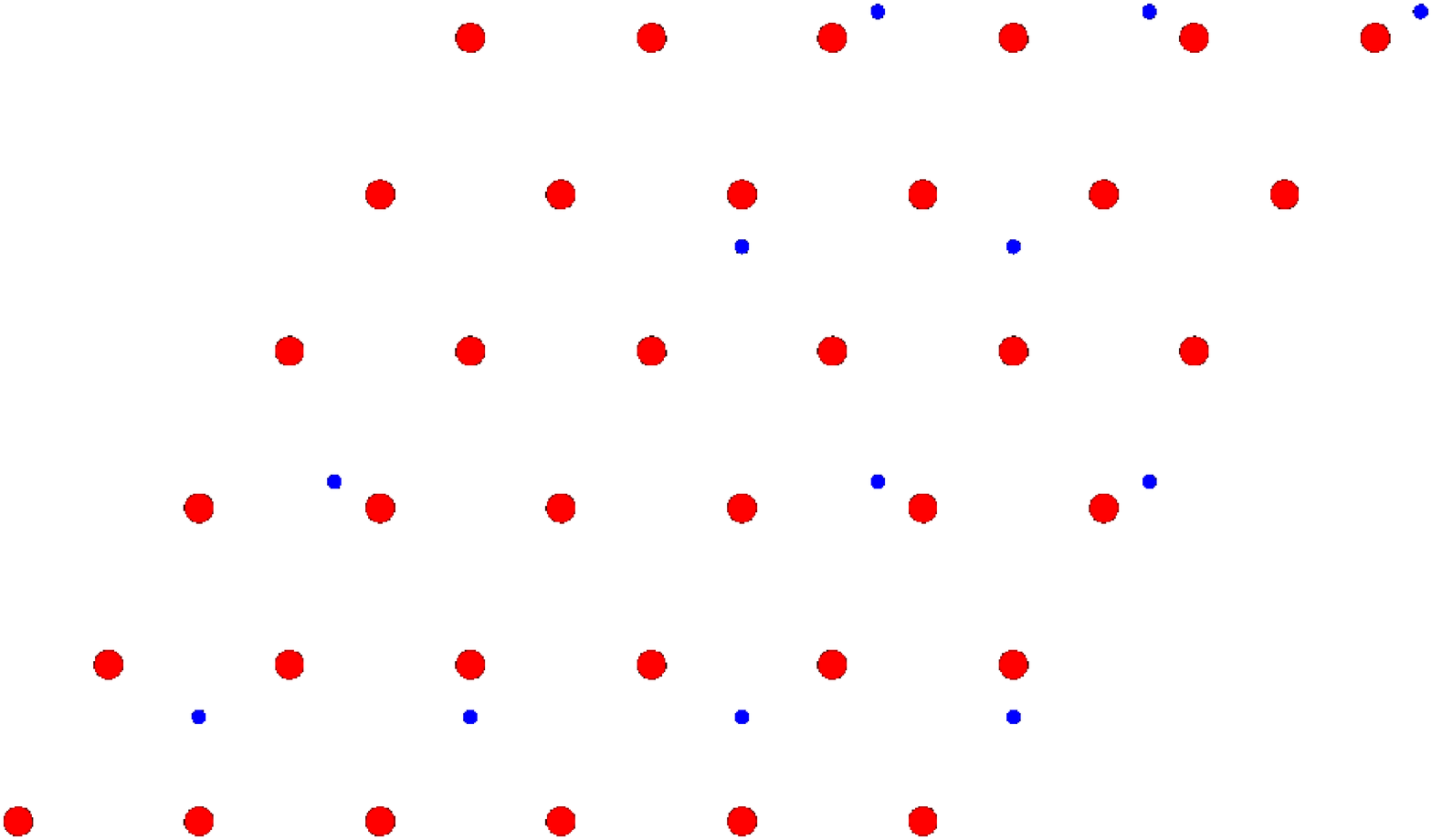}
\makebox[\textwidth][c]{
\begin{tabular}{>{\centering\arraybackslash} m{5cm} | >{\centering\arraybackslash} m{5cm} | >{\centering\arraybackslash} m{5cm}}
{\bf KlimFe-O-Y} & {\bf KlimMixed}
& {\bf KlimFe-Y-O} \\[1mm]
 \includegraphics[width=51mm]{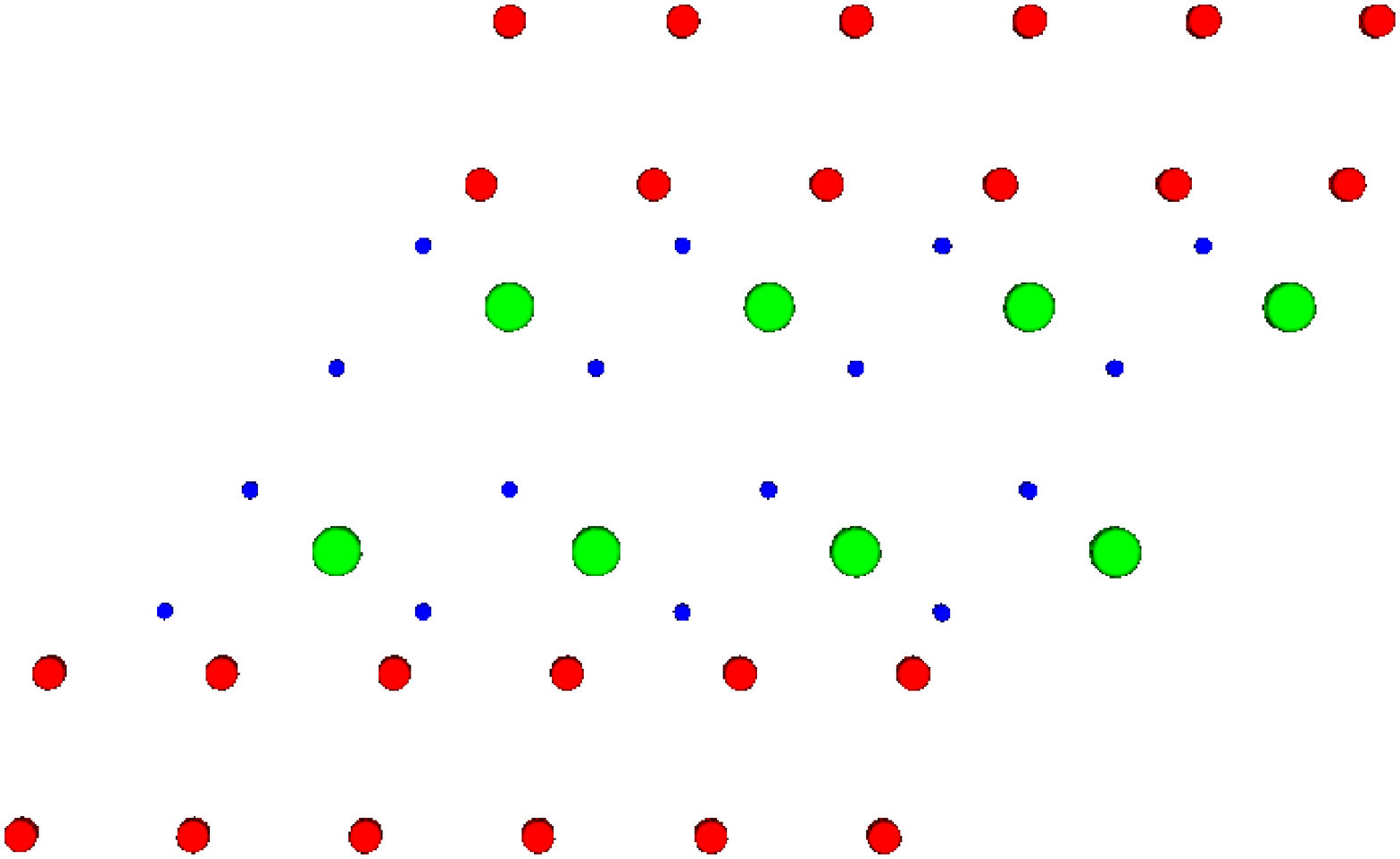} 
& \includegraphics[width=51mm]{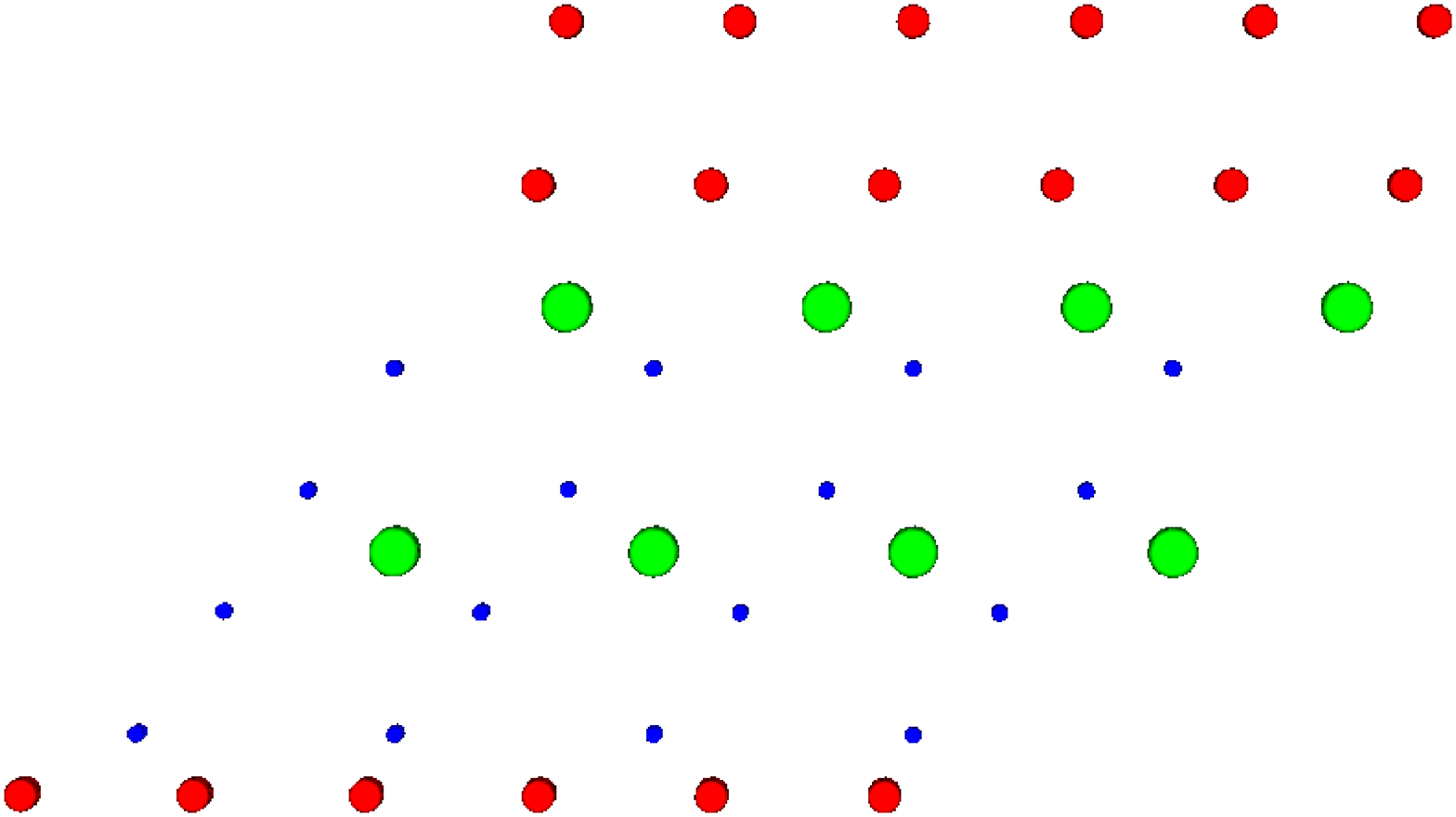}
& \includegraphics[width=51mm]{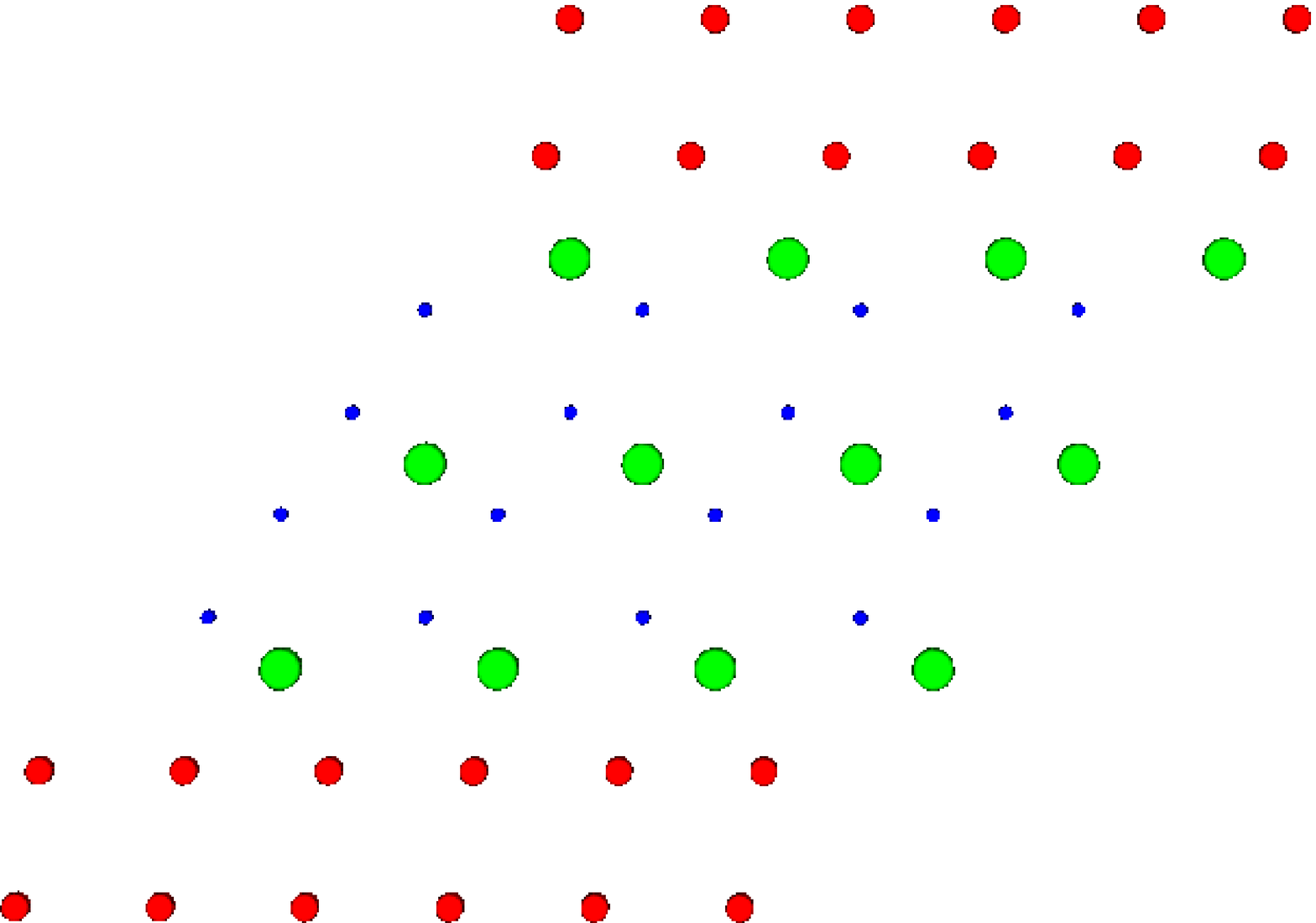}  \\
\end{tabular}}
\caption{\label{fig:KlimSetup} Upper: arrangements of oxygen(blue) and iron (red) atoms in the interfacial plane.  (lower) the three yttria terminations shown viewed
  along the [01$\bar{1}$] axis.}
\end{center}
\end{figure}

BCC iron [111] planes have hexagonal structure, with
directions in this plane initially matched those of
the corresponding pure yttrium oxide cell. The iron atoms at the interface were
initialised to give the smallest interplanar distance, assuming fixed interatomic distances. An
example of how this was achieved in the Fe-O case is shown in
figure \ref{fig:KlimSetup}. The number of iron planes was
determined by requiring that the fit would be similar at top and
bottom of the interface. The setups resulting from periodic boundary
conditions are shown in figure \ref{fig:KlimSetup}.

\subsection{Point Defects}
After the supercells were relaxed, vacancies were formed by removing
iron atoms and SIAs by replacing one iron atom with a dumbbell such
that nearest-neighbour distances were maximised.  Up to four SIAs or
vacancies at the interface were investigated for each case. The low
symmetry ensures that a local minimum energy configuration is found,
but all sites in the interface are inequivalent (see
Fig.\ref{fig:KlimSetup}), so an exhaustive combinatoric search is
impractical.  Thus we selected sites with particularly high or low
atomic density. For example, SIAs were often inserted in large gaps
along, and vacancy sites chosen by removing Fe atoms closest to their
adjacent O or Y. Additionally, the effect of creating a single point
defect in planes further away from the interface was investigated, and
in some cases this led to a more stable structure.

We define the defect energies in the interface relative to defect in
the lattice.  Thus if the defect-free interface calculation has energy
$F$, and the calculation of the interface with M extra Fe atoms has
energy $F_M$, then the formation energy per defect is given by

\[ E = (F_{M}-F-M\mu_{Fe})/M \]

A similar definition is used for vacancies.

\section{Results}



\subsection{Interfacial Energy}

The energy needed to form the stoichiometric interface combinations
from the pure components, and the associated surface energy, $\gamma$,
are given in table \ref{tab:surften}. It can be seen that the
interfaces observed experimentally do indeed have lowest energy. 
\begin{figure}[H]
\begin{center}
\begin{tabular}{|c|ccc|}
\hline
Interface & Formation Energy / eV & $\gamma$ /
eV$\mathrm{\AA^{-2}}$ & $\gamma$ / J$\mathrm{m^{-2}}$\\
\hline 
(100)Mixed & 50.0 & 0.215 & 3.45  \\
KlimFe-Y-O & 65.0  & 0.163 & 2.60 \\ 
KlimMixed & 73.2 & 0.182 & 2.91  \\ \hline
\end{tabular}
\end{center}
\caption{\label{tab:surften}Surface energy, $\gamma$,  for interfaces in stoichiometric cells.
 Note that the ``mixed'' cases are an average of two inequivalent interfaces.}
\end{figure}



We investigated the local electronic density of states projected onto
each atom.  The main feature is that from an electronic point of view,
the interfaces are sharp, with the chemically-expected Y$^{3+}$ and
O$^{2-}$ ions.  The classical method for estimating metal-ionic
interface binding involves the induction of ``image charges'' in the
metal: there is no evidence of any physical image charges underpinning
this calculational trick.  The oxygen valence states lie below Fe
d-band, while the yttrium d-states lie slightly above the iron d-band,
well above the Fermi energy.  Thus there is no metallic or covalent
bonding across the interface.  The sharpness of the interface explains
why the calculations of interface energy converge rapidly with supercell
size, even for non-stoichiometric cells.


\subsection{Point Defects}
\begin{figure}[H]
\begin{center}
\includegraphics[width=110mm]{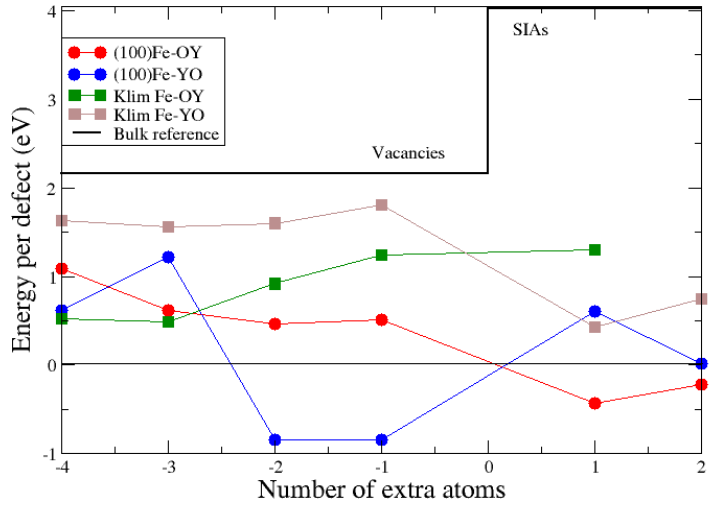}
\end{center}
\caption{\label{fig:ChemPotInt}Defect creation energies for additional
  atoms at the various interfaces (vacancies are shown as negative
  additional atoms). Defect numbers and energies are given relative to
  the ``ideal'' interface. Points represent actual calculations, connecting 
  lines
  are guides to the eye and have no physical significance. 
 The thick black line is the formation energy for
  the equivalent defects in bulk iron. Negative energies indicate that
  the ``ideal'' (100) interface is not the most stable.  Convergence
  is to better than 0.05 eV.  Note that even lower energies for single
  interstitials in the ``Klim'' interfaces were found by relaxation
  from second-layer configurations}
\end{figure}

Figure \ref{fig:ChemPotInt} shows the change in energy as the number
of defects at the interface is altered by adding and removing iron
atoms.  It can be seen that in some cases there is a negative
formation energy for defects in the (100) interface - this is further
evidence of the instability of that interface. In all cases, the
energy costs of vacancy and SIA creation at all interfaces were found
to be consistently less than in bulk iron.  Consequently the ODS
particle acts as a trap for both types of defect.

\begin{figure}[H]
\begin{center}
\makebox[\linewidth][c]{
\begin{tabular}{cc}
\includegraphics[width=90mm]{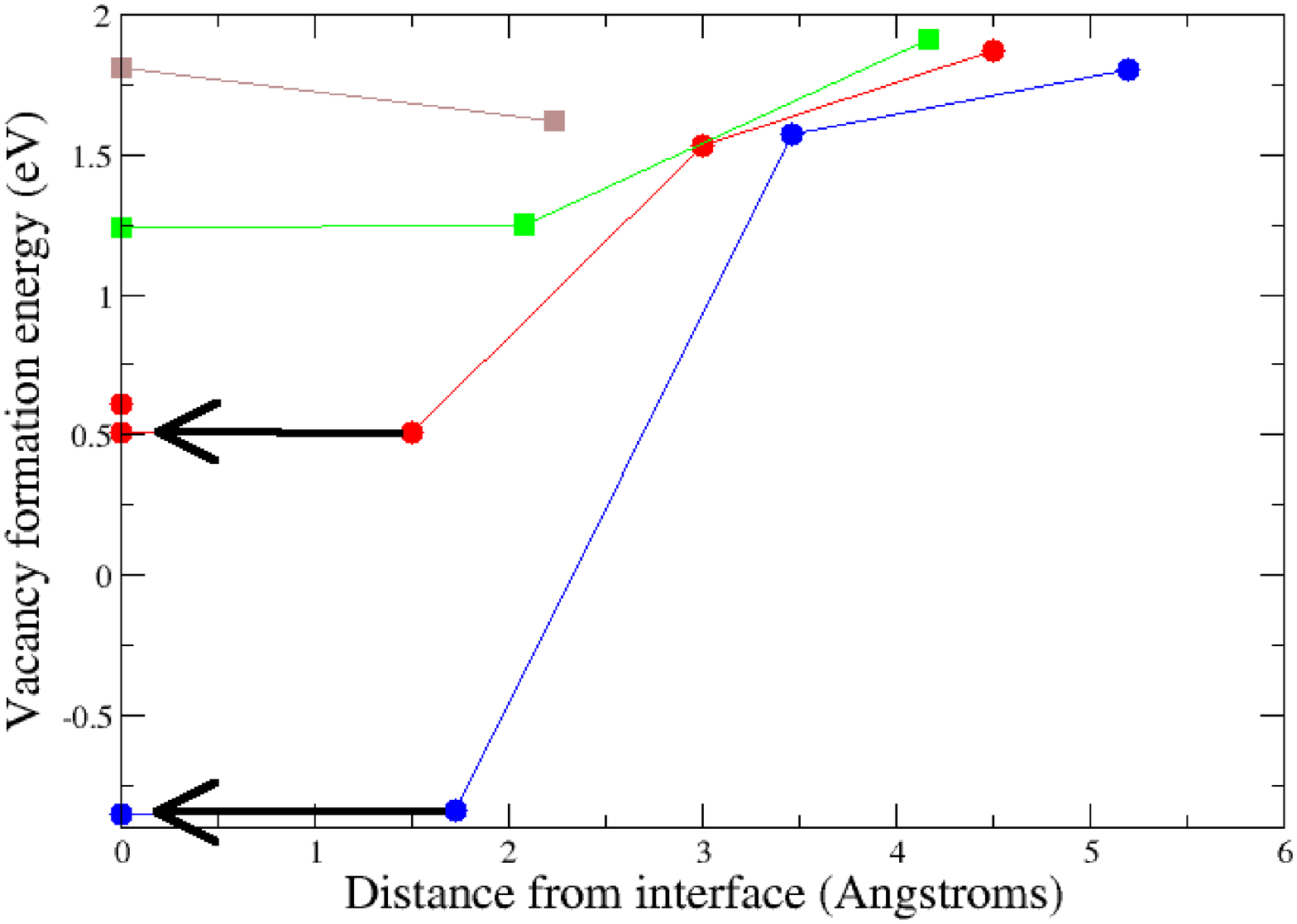}
& \includegraphics[width=90mm]{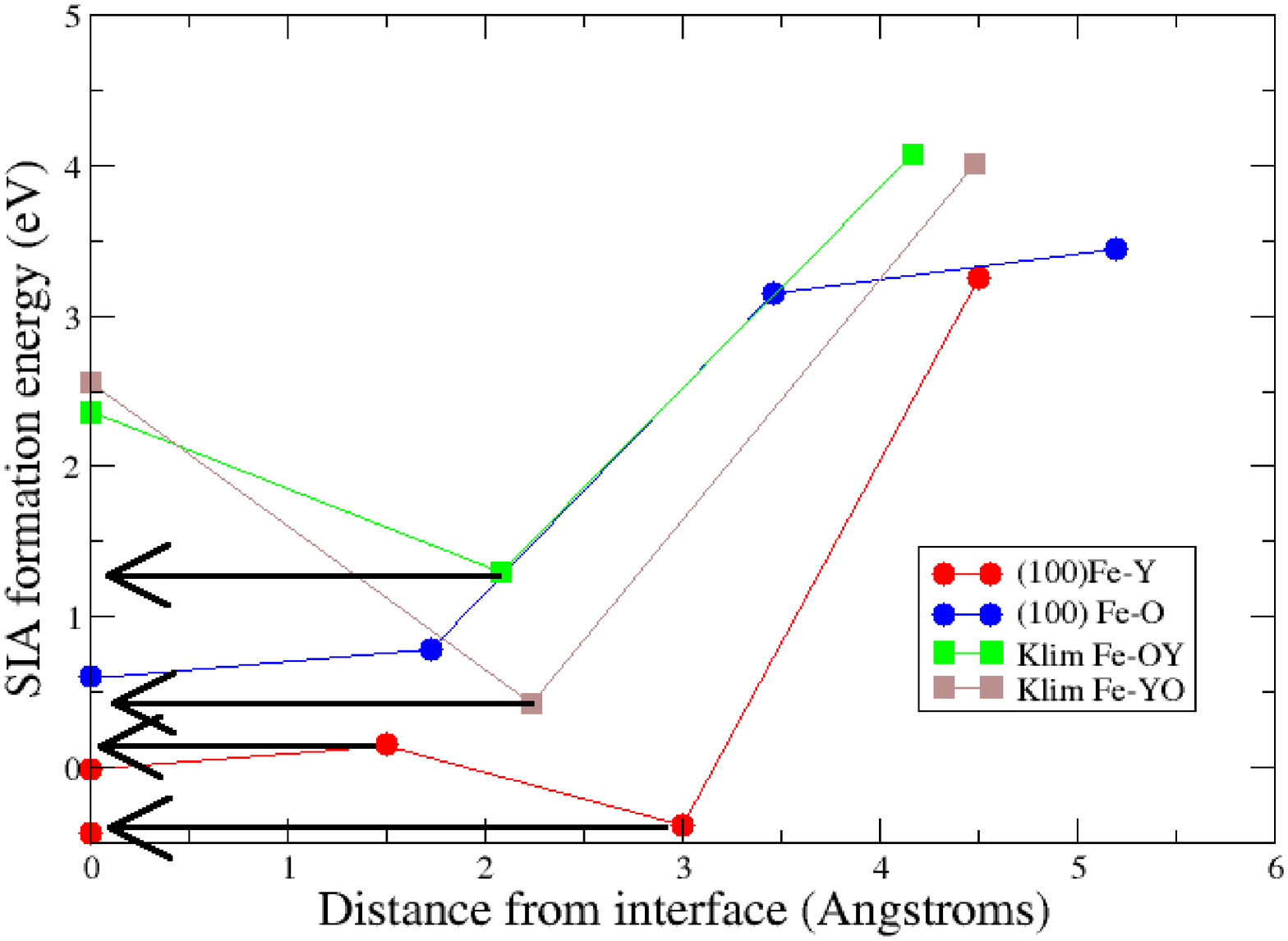} \\
\end{tabular}
}
\end{center}
\caption{\label{fig:VacCost}The calculated variation of vacancy and
  SIA formation energy with increasing distance from the
  interface. Black arrows show where there is a spontaneous relaxation
  into the boundary at a site not previously found.}
\end{figure}

\begin{figure}[H]
\begin{center}
\begin{tabular}{c@{\hspace{5mm}}c@{\hspace{5mm}}c}
\includegraphics[width=35mm]{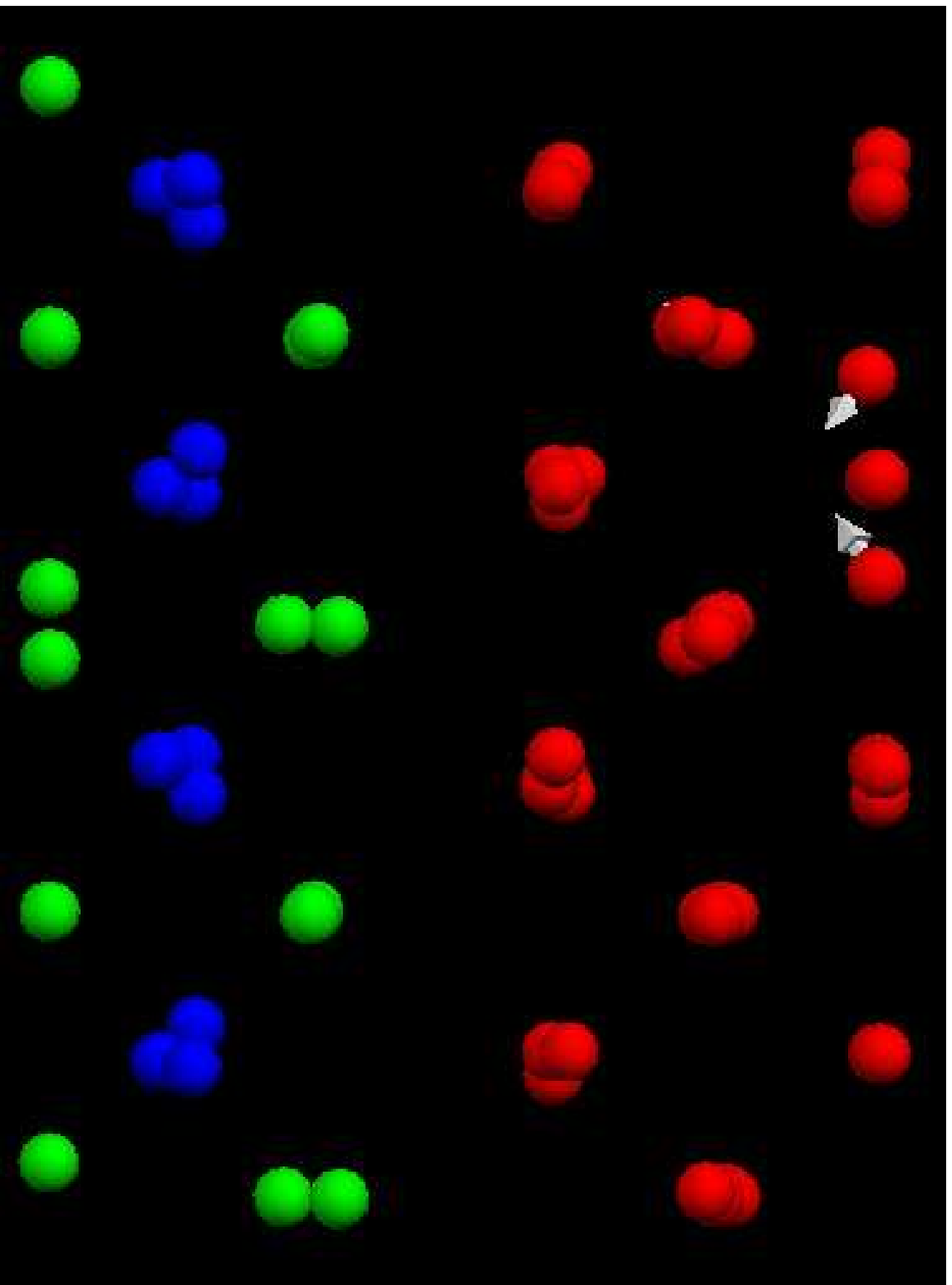} &
\includegraphics[width=35mm]{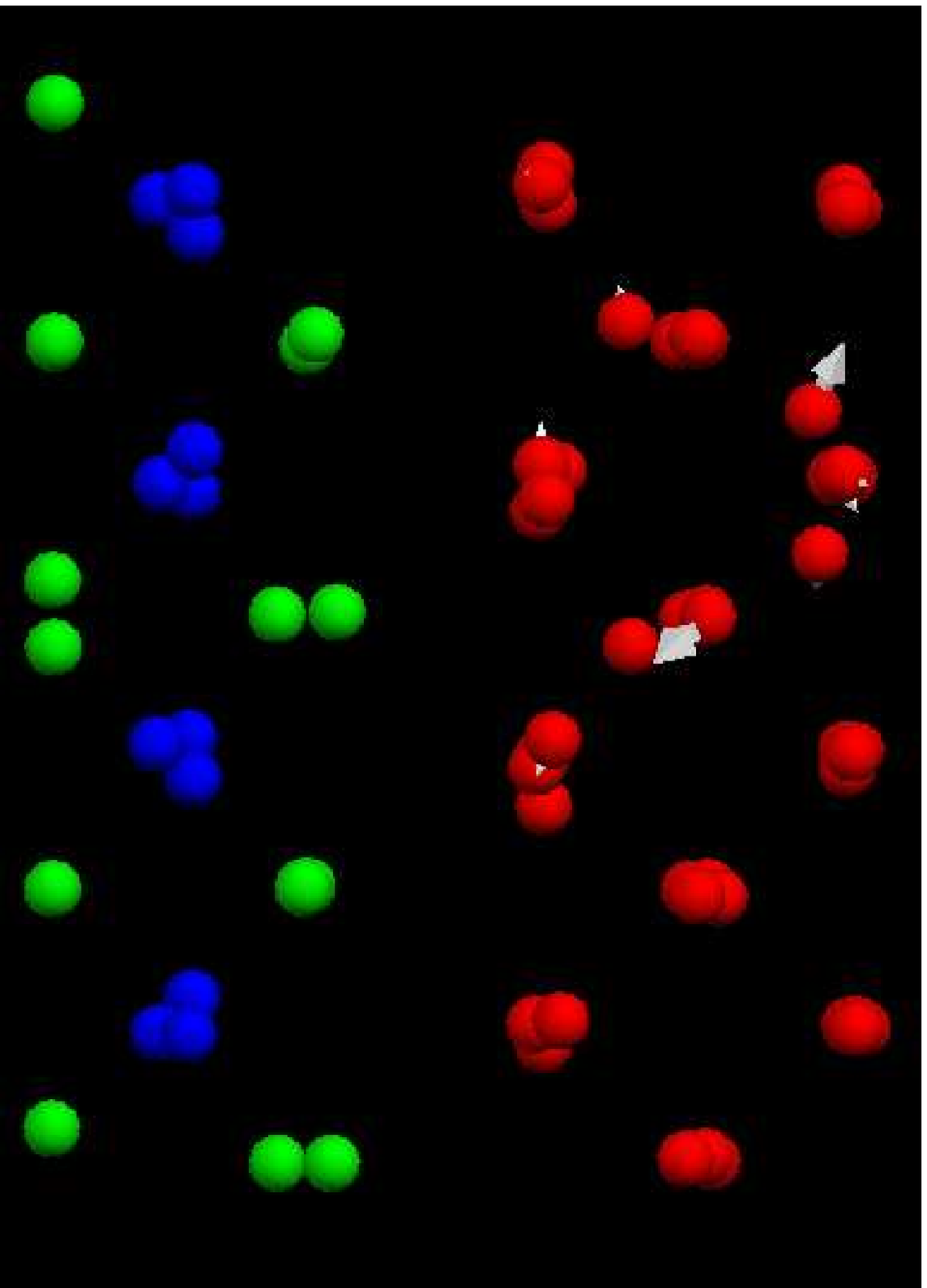} &
\includegraphics[width=35mm]{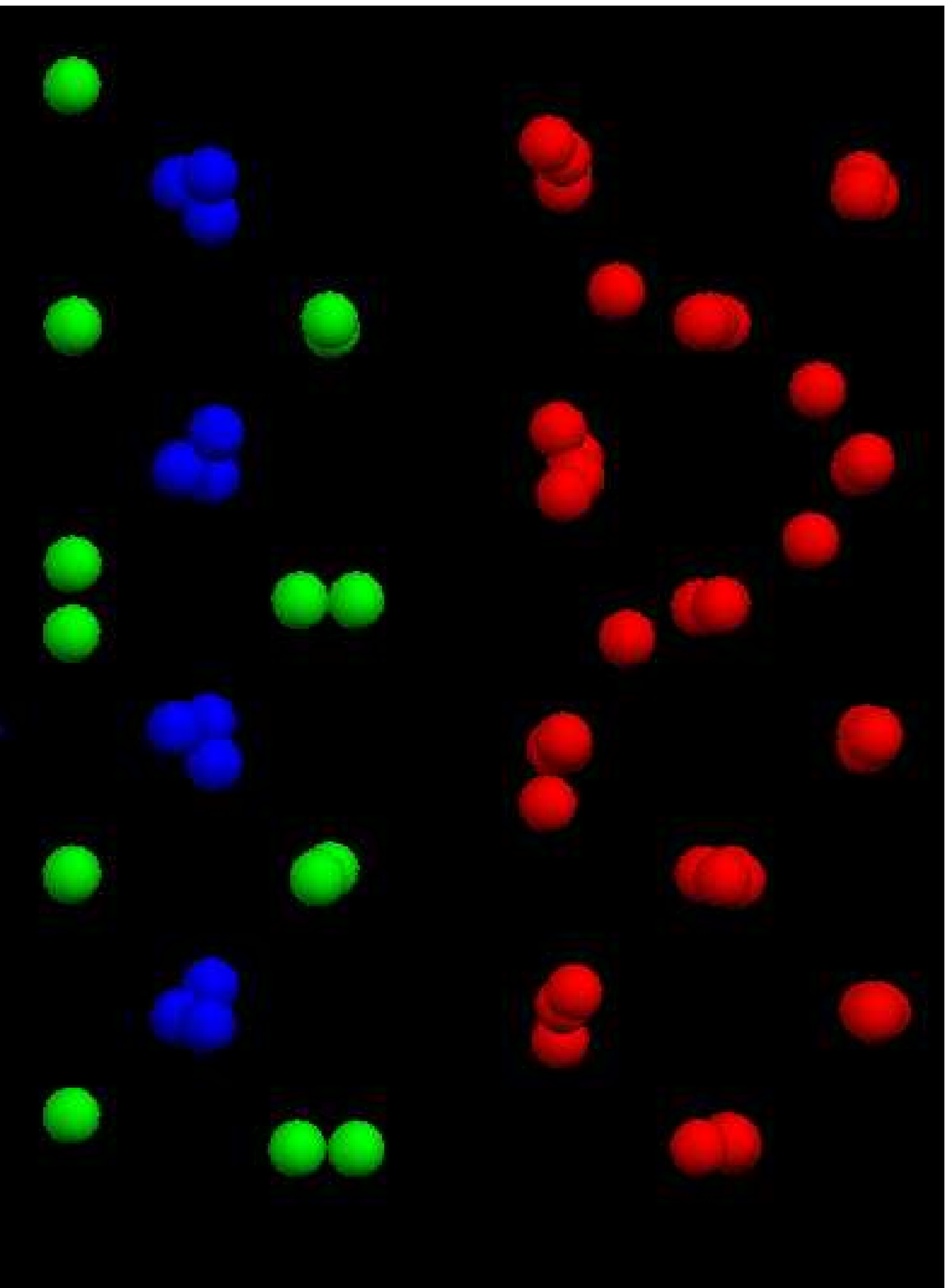} \\
\vspace{5mm}
\includegraphics[width=35mm]{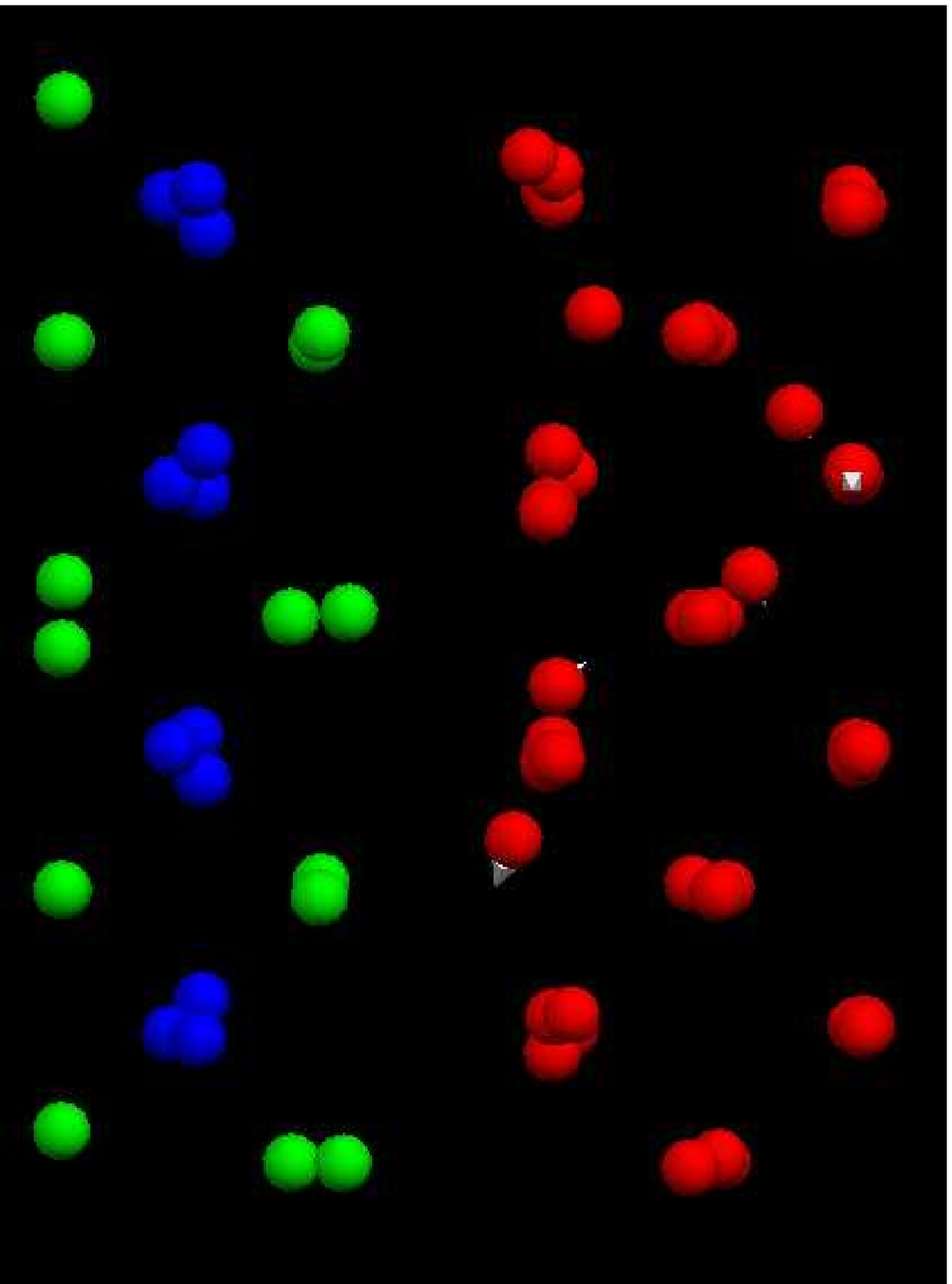} &
\includegraphics[width=35mm]{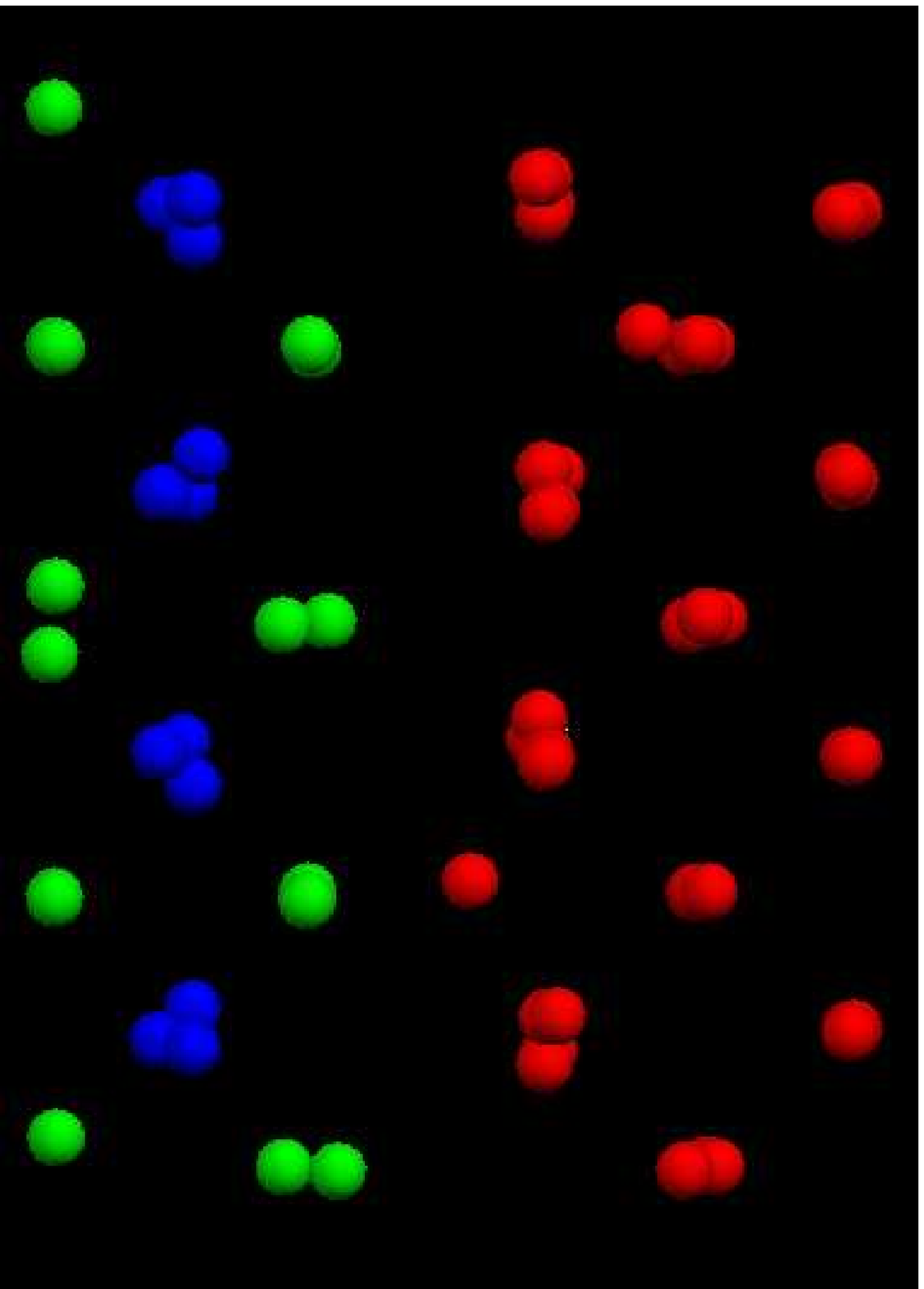} &
\includegraphics[width=35mm]{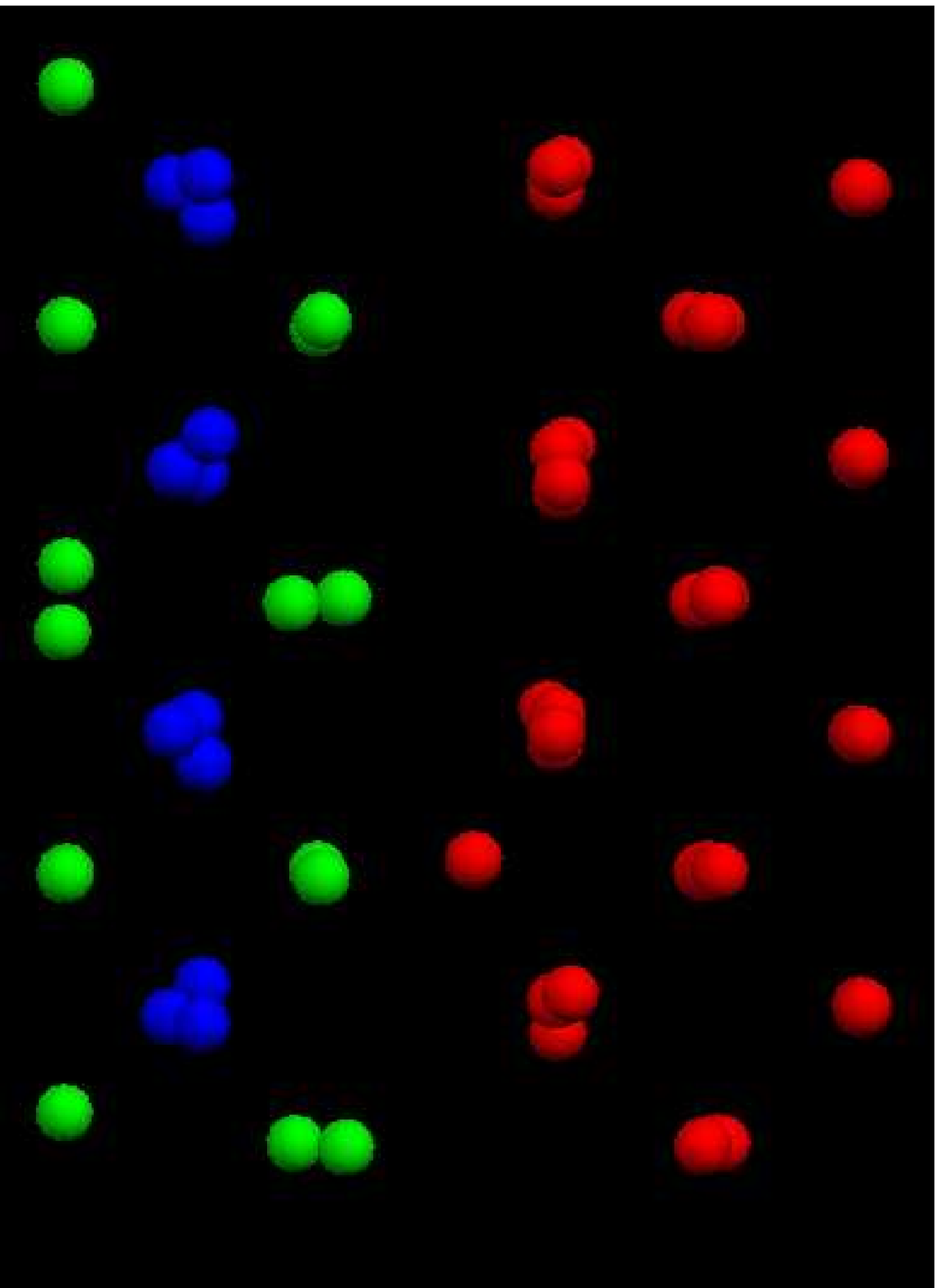} \\
\end{tabular}
\end{center}
\caption{\label{fig:migration} Nearby SIAs were observed to
  migrate towards the (100)Fe-Y interface. The path of migration is
  shown above in stages: An SIA starting towards the top of
  the far-right plane causes a knock on effect in neighbouring atoms that
eventually results in an SIA at the lower part of the interface.}
\end{figure}

Figure \ref{fig:VacCost} shows the point defects energetics at sites
further away from the interface.  Importantly, this is only a subset
of the calculations we performed. In many cases defects that were
created one layer from from the interface spontaneously relaxed into
interface. This was particularly the case with SIAs near iron-yttrium
interfaces as is depicted in figure \ref{fig:migration} for the case
of the (100)Fe-Y interface.  The implication is the that the final
step of capture of defects by ODS particles is essentially
barrierless.  It is also worth noting that, in some cases, the
configuration obtained from interstitial migration has lower energy
than the defect relaxed in the interface.  This implies the existence
of many low energy interface sites for absorption of SIAs and
vacancies, and of easy migration within the interface.  Interface
interstitial configurations set up with (001) dumbbell iron atoms
relaxed to metastable energies of 2.36eV (KlimFe-OY) and 2.56eV
(KlimFe-YO).  These energies (less the stable formation energy 1.30/0.43) 
place a limit on the migration barriers in the interface.

As progressively more atoms are removed from the interface, the
integrity of the layer itself is compromised.  Trivially, removing (or adding) 
16 atom from/to an iron layer (and allowing for a realigning relaxation)
is exactly equivalent to recreating the perfect interface, and would
have zero ``energy per defect''.  15 vacancies would be equivalent to
one interstitial etc.  An interesting conformation is to remove a complete
row of 6 atoms at the KlimFe-O-Y, essentially creating a misfit
dislocation.  This resulted in an energy increase of 3.19 eV (0.53eV/vacancy)

\section{Discussion}

DFT is now well established as a reliable technique for describing
these types of material, so the agreement with experiment for the
perfect materials and superior stability of the Klim interfaces is no
surprise.
The large planar separation, high particle
density and low misfit strain are contributing factors to its
stability. Stoichiometry makes it impossible to define the stable
interface: the Klim-Fe-O-Y termination is certainly more stable than the
average of Klim-Fe-O-O and  Klim-Fe-Y-O  interfaces.  However reconstruction 
 suggests that the double-oxygen layer at the
 KlimFe-O-O is least favoured, probably due to low binding between
the double oxygen layer, so  Klim-Fe-Y-O may be low energy.

Electronic wavefunctions are attenuated
before the second plane of atoms. This suggests that the interface is sharp, 
and electronic effects do not significantly affect the electronic structure 
away from the interface. Thus most
long-range behaviour could be attributed to `physical' rather than
`chemical' effects.

For nuclear applications, the most important results concern the
energetics of point defects.  In all cases both SIA and vacancy
defects are much more stable at the interface than in the bulk iron,
and multiple defects can be accommodated simultaneously.  Moreover, if
a full layer of Fe atoms is created/removed the interface is perfectly
restored, so there is no effect dues to sink strength bias.  Crucially,
unlike grain boundaries, surfaces and other sinks, the oxide
nanoparticle is unaffected by the flow of defects. Incoming vacancies
were not observed to trigger the outward diffusion of Y or O atoms
into the steel. Consequently, ODS particles are perfectly capable of
absorbing defects without contributing to swelling, hardening, creep
etc.  Thus our calculations provide strong evidence in favour of the
catalyst model of the action of ODS particles \citep{radExp}

Furthermore, the energy gradients in figure \ref{fig:VacCost} for the
(100) interfaces and 
numerous barrier-free pathways for inward migration of SIAs near iron-yttrium
interfaces suggest a long range `attraction' of the
interface to point defects\citep{BaVoter}. 

We note that additional alloy elements present in EUROFER97 were not
considered. In fact, an energy-dispersive X-ray study on the
EUROFER97-based ODS steel has shown that the chromium and vanadium
present in the alloy form thin shells around the
nanoparticles\citep{vcro}, presumably attracted by the strain
fields. These elements have their own attraction for point defects, and
consequently could be important in
modifying the interface properties. Unfortunately, this
possibility has had to be neglected in this present investigation.

In sum, our study shows that the ODS-Fe interface has a strong
attraction for vacancy and interstitial defects, and provides strong support for a ``catalytic'' model of point defect removal and consequent radiation resistance.

\section{Acknowledgements}
We wish to acknowledge the use of the EPSRC funded Chemical Database
Service at Daresbury\citep{CDS}, and the UKCP collaboration for supercomputer
time (HECToR).  This work was inspired by the FP7-Getmat project.



\end{document}